\documentclass[aps,pra,twocolumn,showpacs,superscriptaddress,groupedaddress]{revtex4-2}

\usepackage{amsmath,braket,amssymb,natbib,graphicx,float,amsthm,xcolor,outlines,mathrsfs,multirow,hhline}
\usepackage{multirow}
\usepackage{mathtools}
\usepackage{bm}

\DeclareMathOperator{\Tr}{Tr}

\usepackage[english]{babel}
\usepackage[utf8x]{inputenc}
\usepackage[T1]{fontenc}
\usepackage{listings}
\usepackage[colorinlistoftodos]{todonotes}
\usepackage[colorlinks=true, allcolors=blue]{hyperref}
\begin{document}
\title{Genuine non-Gaussian entanglement of light and quantum coherence for an atom from noisy multiphoton spin-boson interactions}

\author{Pradip Laha}
\email{plaha@uni-mainz.de}
\affiliation{Institute of Physics, Johannes-Gutenberg University of Mainz,  Staudingerweg 7, 55128 Mainz, Germany}
\author{P. A. Ameen Yasir}
\email{apooliab@uni-mainz.de}
\affiliation{Institute of Physics, Johannes-Gutenberg University of Mainz,  Staudingerweg 7, 55128 Mainz, Germany}
\author{Peter van Loock}
\email{loock@uni-mainz.de}
\affiliation{Institute of Physics, Johannes-Gutenberg University of Mainz,  Staudingerweg 7, 55128 Mainz, Germany}


\begin{abstract}
Harnessing entanglement and quantum coherence plays a central role in advancing quantum technologies. In quantum optical light-atom platforms, these two fundamental resources are often associated with a Jaynes-Cummings model description describing the coherent exchange of a photon between an optical resonator mode and a two-level spin. In a generic nonlinear spin-boson system, more photons and more modes will take part in the interactions. Here we consider such a generalization -- the two-mode multiphoton Jaynes-Cummings (MPJC) model. We demonstrate how entanglement and quantum coherence can be optimally generated and subsequently manipulated in experimentally accessible parameter regimes. A detailed comparative analysis of this model reveals that nonlinearities within the MPJC interactions produce genuinely non-Gaussian entanglement, devoid of Gaussian contributions, from noisy resources. More specifically, strong coherent sources may be replaced by weaker, incoherent ones, significantly reducing the resource overhead, though at the expense of reduced efficiency. At the same time, increasing the multiphoton order of the MPJC interactions expedites the entanglement generation process, thus rendering the whole generation scheme again more efficient and robust. We further explore the use of additional dispersive spin-boson interactions and Kerr nonlinearities in order to create spin coherence solely from incoherent sources and to enhance the quantum correlations, respectively. As for the latter, somewhat unexpectedly,   there is not necessarily an increase in quantum correlations due to the augmented nonlinearity. Towards possible applications of the MPJC model, we show how, with appropriately chosen experimental parameters, we can engineer arbitrary NOON states as well as the tripartite W state.
\end{abstract}

\maketitle


\section{\label{sec:intro}Introduction}
Quantum entanglement~\cite{Ekert_1998,horodecki_RMP_2009} and quantum coherence~\cite{plenio_coherence_rmp_17,Streltsov_njp_2018},  in the context of both discrete variable (DV)~\cite{nielsen_chuang_2010} as well as continuous variable (CV)~\cite{Braunstein_RMP_2005} quantum systems, are two of the most prominent resources available for modern quantum technology, such as quantum computation~\cite{nielsen_chuang_2010}, quantum key distribution~\cite{Ahonen_PRA_2008,Xu_RMP_2020}, quantum cryptography~\cite{Gisin_RMP_2002}, and quantum sensing and quantum metrology~\cite{Giovannetti_PRL_2006,Degen_RMP_2017}. Thus, on-demand generation and efficient manipulation of these resources have been the subject of active research in the past decades (see, for instance, Refs.~\cite{Raimond_RMP_2001,Li_Sci_rep_2012,Streltsov_PRL_2015,Wang_PRL_2017,Zhao_PRA_2021,Piccione_PRA_2022,shiraishi_2023,kondra_coherence_2023}).
The Jaynes-Cummings model (JCM)~\cite{jaynes_cummings_1963}, capturing the inherently nonlinear spin-boson interaction between a \textcolor{blue}{CV} bosonic mode (oscillator) and a \textcolor{blue}{DV} two-level system (spin/qubit) in the rotating wave approximation, is a cornerstone in quantum physics, playing a pivotal role as a test bed for exploring quantum optics and phenomena in quantum information science~\cite{Shore_JCM_1993}. Experimentally, this model has been successfully implemented across various quantum platforms currently available. Examples include cavity QED~\cite{Rempe_PRL_1987,Brune_PRL_1996}, circuit QED~\cite{Deppe_nat_phys_2008,Fink_nature_2008}, ion traps~\cite{Leibfried_RMP_2003,Lara_PRA_2005},  and others~\cite{Dora_PRL_2009,Basset_PRB_2013,Lee_PRA_2017}. 

Given its immense importance, this foundational model has undergone extensive generalization over the years to accommodate diverse forms that accurately capture the quantum features exhibited by specific, realistic physical systems (see Ref.~\cite{Larson_JCM_2021} for a comprehensive review).
A noteworthy theoretical extension developed over the years involves the multiphoton Jaynes-Cummings (MPJC) model. For a single bosonic mode characterized by the bosonic creation and annihilation operators $a^\dagger$ and $a$, respectively, the interaction Hamiltonian for the $m-$photon JCM is expressed as $g(a^m\sigma_+ + a^{\dagger\, m} \sigma_-)$, where $\sigma_+ = \ket{e}\bra{g}$ and $\sigma_- = \ket{g}\bra{e}$ denote the raising and lowering operators of the two-level system with levels $\ket{g}$ and $\ket{e}$, respectively. The parameter $g$ represents the strength of the  $m-$photon interaction, and $m=1, 2, \cdots$ signifies the degree of the nonlinear processes. Note that the case $m=1$ corresponds to the standard JCM. Theoretically, an array of quantum mechanical phenomena intrinsic to the MPJC interaction has been investigated comprehensively over the years, encompassing aspects ranging from field statistics to squeezing dynamics (see, e.g., Refs.~\cite{SUKUMAR_pla_1981,surendra_pra_1982,SHUMOVSKY_pla_1987,Kien_PRA_1988,LuHuai_Chin_phys_2000,El_Orany_job_2003,El_Orany_jpa_2004,Villas_PRL_2019}).

A natural extension of the single-mode $m$-photon JCM involves the concurrent interaction of two bosonic modes with a single two-level system. In particular, we are primarily interested in the entanglement dynamics between the two CV modes, denoted as bosonic entanglement, and the coherence manifested in the DV qubit, with a specific emphasis on how these phenomena evolve under varying degrees of nonlinearity. It is pertinent to mention that while diverse generalizations of the $m$-photon JCM involving two bosonic modes and a single two-level system are documented in the literature (see, for example,~\cite{Cardimona_PRA_1991,El_Orany_oc_2004,singh_ijtp_2019,Alushi_PRXQuantum_2023}), the specific tripartite system under consideration for our present investigation has not been hitherto analyzed.

Recent progress in experimental platforms, including generic spin-boson models, ion traps, and superconducting circuits, indicates that exploring such nonlinear spin-boson interactions is on the horizon, achievable with upcoming quantum technology. In this work, we focus specifically on the rapidly advancing field of spin-boson systems. The study of spin-boson Hamiltonians and their applications spans various quantum science and technology domains, including quantum simulation~\cite{Puebla_PRA_2017,Puebla_symm_2019} and quantum computation~\cite{Miessen_PRRR_2021}, quantum thermodynamics~\cite{Rivas_PRL_2020}, and phase transitions~\cite{vojt_phil_mag_2006}, just to name a few. The archetypal spin-boson model involves a spin interacting with an environment composed of a continuum of bosonic modes.  Despite the seemingly limited control, a transformative method has been previously established, effectively mapping the spin-boson dynamics to that of a tunable MPJC model undergoing dissipation~\cite{Puebla_symm_2019}. This serves as the foundation for our current exploration.

The generation of quantum correlations in any quantum system is highly sensitive to the initial state preparation. Conventionally,  a strong coherent external source is necessary for the entanglement and the coherence to materialize. However, recent studies have shown that low incoherent energy has the remarkable ability to produce genuine nonclassical correlations in nonlinear quantum systems~\cite{marek_nonclass_pra_16,slodicka_oe_16,laha_thermally_2022,laha_non-gaussian_2022,cusumano_structured_2003}, thereby circumventing the need of such highly coherent resources. Although the nonclassical correlations generated by incoherent resources are generally lower than their coherent counterparts, previous research has demonstrated that this efficiency challenge can be addressed through protocols such as entanglement and coherence distillation~\cite{ent_dist_PRA_2018, shiraishi_2023}, allowing for arbitrary amplification of such quantum correlations. Furthermore, a comprehensive comparative analysis of the impact of coherent versus incoherent state preparation on the entanglement and coherence dynamics of a generic nonlinear quantum system, as explored in this study, is currently lacking. We contend that this article significantly addresses this gap.

Among other results, we report that (i) while bosonic entanglement can be readily generated using only incoherent noisy resources, the nonlinearity within the MPJC proves to be insufficient in generating coherence in the two-level system and it requires additional nonlinear dispersive spin-boson interactions, (ii) the generated entanglement exhibits genuinely non-Gaussian characteristics with no Gaussian contributions, (iii) increasing the order of the MPJC interactions (denoted as $m$) accelerates the entanglement generation process, thus rendering the whole generation scheme more efficient and robust, and (iv) contrary to intuition, augmenting nonlinearity in the governing Hamiltonian by introducing extra Kerr nonlinearities does not necessarily amplify these quantum correlations. To place these results within a broader experimental framework, we provide a detailed account of environmental-induced effects on the system's dynamics. 

This article's novelty lies in (i) the choice of a tripartite nonlinear quantum model, which is generic, suggesting that results applicable to many important models can be obtained by appropriately choosing $m$, (ii) the extension of the model to incorporate experimentally relevant nonlinear Kerr effects, (iii) the comprehensive comparative study of the distinct dynamics of bosonic entanglement and spin coherence generated using incoherent and coherent resources, (iv) the significant generalization of results on spin coherence previously presented in~\cite{laha_rabi_2023}, and (v) the potential application in engineering specific target non-Gaussian entangled states such as the bipartite NOON states and the tripartite W state~\cite{Dur2000} which are important resources in optical quantum information science~\cite{Dowling_2008,Bergmann_PRA_2016} where the former state has been widely used in quantum metrology~\cite{Giovannetti_nat_photon_2011}, and quantum lithography~\cite{Boto_PRL_2000}.

The rest of the paper is arranged as follows: In Section~\ref{sec:model}, we give a very brief description of the model, while in Section~\ref{sec:results} we outline our main results in the absence of environmental effects. In Section~\ref{sec:results_open} we detail our results on the role of system-environment coupling on the dynamics in the Markovian limit. We conclude with a brief discussion and point towards avenues for further research in Section~\ref{sec:conc}. Additionally, a set of appendices augmenting the results presented in this paper is included at the end.

\begin{figure*}[htbp]
\centering
\includegraphics[scale=0.43]{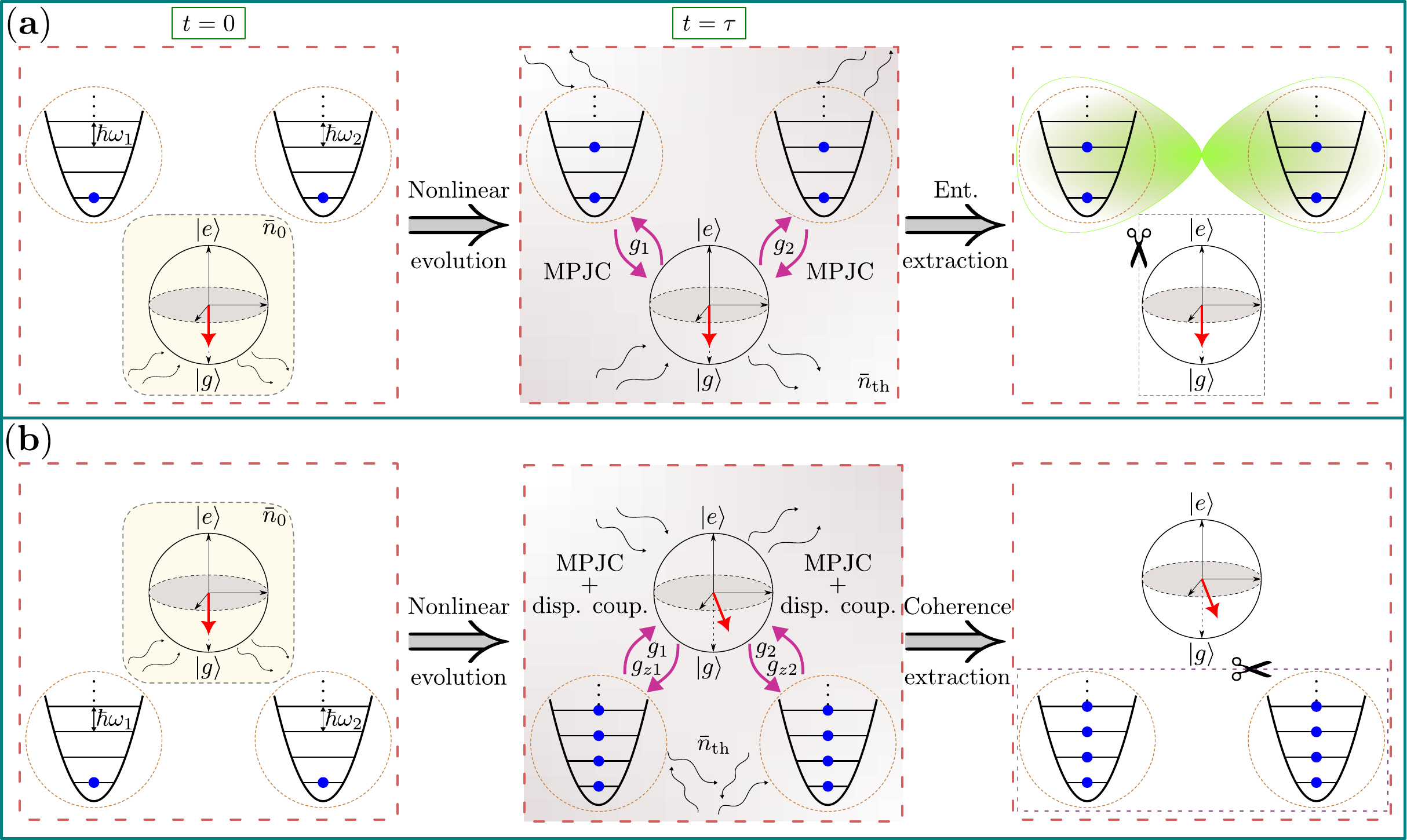}
\caption{The schematic representation illustrates the emergence of (a) bosonic entanglement and (b) spin coherence in a generic nonlinear spin-boson model, effectively comprising two bosonic modes and a single spin, initially excited to an incoherent noisy state (depicted by red arrows) through interaction with a hot bath. The bosonic modes are prepared in the ground states. The tripartite system subsequently undergoes evolution under various nonlinear interactions. In the former case, the initially unentangled bosonic modes become entangled solely through the two-mode multiphoton Jaynes-Cummings (MPJC) interactions with respective coupling strengths $g_1$ and $g_2$. The entanglement is discerned by tracing out the spin degrees of freedom from the time-evolved tripartite state at an optimal time $\tau$. The resulting entanglement is {\em genuinely} non-Gaussian (see main text) and persists even in the presence of various decoherence and dephasing channels (depicted by various curvy arrows), including incoherence in the initial bosonic modes. The common bath temperature, denoted by $\bar{n}_{\text{th}}$, characterizes the thermal environment interacting with all subsystems.  Further insights into the implications of utilizing an incoherent noisy spin state over a coherent one can be found in the main text. Meanwhile, for the latter case starting from such a noisy incoherent spin, the nonlinearities within MPJC interactions are found to be inadequate in achieving spin coherence. Additional nonlinear dispersive spin-boson interactions (with respective coupling strengths $g_{z_1}$ and $g_{z_2}$) become imperative for this purpose.}
\label{Fig_model}
\end{figure*}


\section{\label{sec:model}Two-mode multi-photon JCM}
The results presented in the following are independent of the choice of any specific quantum platform. Nevertheless, as eluded in the preceding section, we will be focusing specifically on spin-boson models in which a two-level system interacts with a large, often infinite, number of bosonic modes, which constitute the environment. Under appropriate parameter regimes, however, such a generic model can be mapped onto an effective tripartite nonlinear quantum system comprising a single two-level system simultaneously interacting with two independent bosonic modes through the $m$-photon Jaynes-Cummings interactions (see Ref.\cite{Puebla_symm_2019} for details). The Hamiltonian (setting  $\hbar=1$) governing such an effective system can be written as
\begin{align} 
    H =  \frac{\omega_0}{2}\sigma_z  + \sum_{i=1}^2 \left[\omega_i a_i^{\dagger}a_i +  g_i\left(a_i^m\sigma_+ + a_i^{\dagger\, m} \sigma_-\right)\right].
    \label{eqn_mpjcm}
\end{align}
Here, $\sigma_z = \ket{e}\bra{e}-\ket{g}\bra{g}$ is the usual Pauli spin operator of the two-level system with levels $\ket{g}$ and $\ket{e}$, respectively, and $\omega_0$ denotes the energy difference between the two levels.  The two bosonic modes are described by the operators $a_i, a_i^\dagger$, which satisfy the standard commutation relation $[a_i,\, a_j^\dagger] = \delta_{ij}$ with $i,\, j = 1,\, 2$,  and $\omega_i$ is the frequency of the bosonic mode $i$. The parameter $g_i$ signifies the strength of the nonlinear MPJC interaction, with the integer $m$ tracking the $m$-photon process. The two detuning parameters are $\Delta_i = \omega_0 - m\omega_i$. 
Throughout this work, we assume $\omega_1=\omega_2$ for simplicity, so that we have a single detuning parameter $\Delta=\Delta_1=\Delta_2$.

A brief derivation of the Hamiltonian in Eq.\eqref{eqn_mpjcm} from the fundamental spin-boson interaction can be found in Appendix~\ref{sec:derivation}. We note in passing that the tripartite JC model, corresponding to $m=1$, has been extensively studied in the past~\cite{laha_thermally_2022,Abdalla_opt_commun_2002}. 

The entanglement between the two bosonic modes is characterized in terms of the logarithmic negativity $\mathcal{L}\equiv \log_2||\rho^{\rm PT}||$, where $\rho^{\rm PT}$ denotes the partial transpose of the two-mode density matrix $\rho$ concerning one of the modes and $\vert\vert \rho^{\rm PT} \vert\vert$ denotes the trace norm~\cite{vidal,plenio_logarithmic_2005}. Alternatively, $\mathcal{L}$ can also be obtained from the negative eigenvalues of $\rho^{\rm PT}$, as given by $\mathcal{L} = \log_2\left[1+ 2 \vert\sum_{\lambda_i<0}\lambda_i \vert \right]=  \log_2\left[1+(|\lambda_{i}|-\lambda_{i})\right]$
where $\lambda_i$ are all of the eigenvalues. 
On the other hand, for the two-level system described by the density matrix $\rho_{\text{s}}$, we employ the coherence monotone $\mathcal{C}=S\left(\text{diag}\left(\rho_{\text{s}}\right)\right)-S\left(\rho_{\text{s}}\right)\,$ as a measure of spin coherence~\cite{plenio_coherence_prl_14}, where $S(\rho_{\text{s}})=-\Tr(\rho_{\text{s}}\log\rho_{\text{s}})$ is the standard von Neumann entropy.

\section{\label{sec:results}Closed system dynamics} We start the analysis under the ideal assumption of unitary dynamics, where dissipation effects are considered negligible and can be ignored. This assumption, however, will be revisited in the subsequent section. The fundamental framework of this article is schematically depicted in Fig.~\ref{Fig_model}.

\subsection{\label{subsec:1}Coherent versus incoherent initial spin}
First we consider that there is no entanglement between the two bosonic modes and that they are initially prepared in their respective ground states $\ket{0}$.
Now, if the two-level system is assumed to be initially  prepared in a generic coherent superposition state $\cos\phi\ket{g}+\sin\phi\ket{e}$, the state of the tripartite quantum system before the temporal evolution has the form
\begin{align}
    \ket{\psi(0)}^{\text{sup}} = \cos\phi\ket{g, 0, 0} + \sin\phi\ket{e, 0, 0},
    \label{psi0_osc_vac}
\end{align}
where the notation is self-evident. The state vector at a later time $t$ for such an initial state can be written as
\begin{align}
  \ket{\psi(t)}^{\text{sup}} = x_1\ket{g, 0, 0} &+ x_2\ket{e, 0, 0} \nonumber \\
  &+ x_3 \ket{g, m, 0} + x_4 \ket{g, 0, m},
  \label{eqn_psi_full_sup_qbt_vac_osc}
\end{align}
where the time-dependent coefficients are found to be
\begin{subequations}
\begin{align}
    x_1 &= \cos\phi \,e^{i\omega_0 t/2}, \\
    x_2 &= \sin\phi\cos\left(\sqrt{m!(g_1^2+g_2^2)}\,t\right)  \,e^{-i\omega_0 t/2} , \\
    x_3 &= \frac{-ig_1}{\sqrt{g_1^2+g_2^2}}  \sin\phi  \sin\left(\sqrt{m!(g_1^2+g_2^2)}\,t\right)  \,e^{-i\omega_0 t/2}, \\
    x_4 &=  \frac{-ig_2}{\sqrt{g_1^2+g_2^2}} \sin\phi  \sin\left(\sqrt{m!(g_1^2+g_2^2)}\,t\right)  \,e^{-i\omega_0 t/2}.
\end{align}
\end{subequations}
Here, we assume a perfect resonance scenario, that is, $\Delta=0$, for simplicity. The general solution with nonzero detuning and its importance on the dynamics, however, can be found in Appendix~\ref{sec:coupled_eqns}. 

In contrast, if we consider the case where the spin is prepared in some generic incoherent thermal state $p_g \ket{g}\bra{g} + p_e \ket{e}\bra{e}$, the initial state of the full system can be expressed as
\begin{align}
    \rho^{\text{th}}(0) = p_g \ket{g, 0, 0}\bra{g, 0, 0} + p_e \ket{e, 0, 0}\bra{e, 0,0}.
    \label{eqn_rho0_th}
\end{align}
The total density matrix at a later time $t$ is found to be (noting that $p_e=\sin^2\phi$)
\begin{align}
  \rho^{\text{th}}(t) &= |x_1|^2 \ket{g,0,0} \bra{g,0,0} + |x_2|^2 \ket{e,0,0}  \bra{e,0,0} \nonumber \\
          &+ |x_3|^2 \ket{g,m,0} \bra{g,m,0} + |x_4|^2 \ket{g,0,m}  \bra{g,0,m} \nonumber \\
          &+ x_2x_3^* \ket{e,0,0}\bra{g,m,0} + x_2^*x_3 \ket{g,m,0} \bra{e,0,0}   \nonumber \\
          &+ x_2x_4^* \ket{e,0,0}\bra{g,0,m} + x_2^*x_4 \ket{g,0,m} \bra{e,0,0}   \nonumber \\
          &+ x_3x_4^* \ket{g,m,0}\bra{g,0,m} + x_3^*x_4 \ket{g,0,m} \bra{g,m,0}.
    \label{eqn_rho_full_th_qbt_vac_osc}
\end{align}
\\

\noindent{\bf Bosonic entanglement:} In both cases, to extract the entanglement between the two bosonic modes, we should first obtain the corresponding reduced two-mode bosonic density matrices $\rho_{\text{b}}(t)$ by tracing out the spin degrees of freedom from Eqs.~\eqref{eqn_psi_full_sup_qbt_vac_osc} and \eqref{eqn_rho_full_th_qbt_vac_osc}, respectively. Given that only two Fock states ($\ket{0}$ and $\ket{m}$, respectively) contribute to the dynamics for each bosonic mode, we can effectively express the two-mode states in the basis $\ket{00}$, $\ket{0m}$, $\ket{m0}$, and $\ket{mm}$.
For the initial superposition spin state, we obtain
\begin{align}
  \rho_{\text{b}}^{\text{sup}}(t) = 
    \begin{pmatrix}
        |x_1|^2 + |x_2|^2 & x_1x_4^* & x_1x_3^* & 0 \\
        x_1^* x_4 & |x_4|^2 & x_3^*x_4 & 0 \\
        x_1^* x_3 & x_3x_4^* & |x_3|^2 & 0 \\
        0 & 0 & 0 & 0
    \end{pmatrix},
    \label{eqn_rho_12_q_sup}
\end{align}
while for the initial incoherent spin, we get
\begin{align}
  \rho_{\text{b}}^{\text{th}}(t) = 
    \begin{pmatrix}
        |x_1|^2 + |x_2|^2 & 0 & 0 & 0 \\
         0& |x_4|^2 & x_3^*x_4 & 0 \\
         0 & x_3x_4^* & |x_3|^2 & 0 \\
        0 & 0 & 0 & 0
    \end{pmatrix}.
    \label{eqn_rho_12_q_th}
\end{align}

%
\begin{figure*}[ht!]
    \centering
    \includegraphics{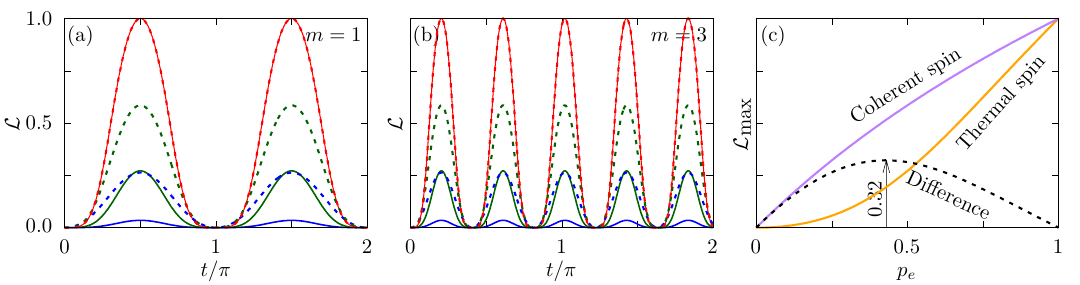}
    \vspace{-4ex}
    \caption{Unitary evolution of {\em genuinely} non-Gaussian bosonic entanglement, characterized in terms of logarithmic negativity $\mathcal{L}$, for the two-mode $m$-photon JCM when both the bosonic modes are initially prepared in their respective ground states for (a) $m=1$ and (b) $m=3$, respectively. The solid (dashed) lines correspond to the initial incoherent noisy thermal (coherent superposition) spin state. The blue, green, and the red curves correspond to $p_e = \sin^2\phi = 0.2$, 0.5, and 1, respectively. We set $g = g_1 = g_2 = \frac{1}{\sqrt{2}}$. Interestingly, for both initial states, the entanglement continues to exhibit perfect oscillatory dynamics with identical Rabi frequency with  $p_e$ controlling the maximum achievable bosonic entanglement ($\mathcal{L}_{\text{max}}$) which is different for both the initial spin states with similarly chosen system parameters (unless, of course, $p_e=0$ or 1). As $m$ is increased to higher values, the frequency of oscillations increases by a factor $\sqrt{m!}$ for fixed $g$. The amplitude of oscillation remains the same suggesting that by considering higher-order JC interaction, maximal entanglement can be reached much faster and more frequently. This becomes particularly advantageous when accounting for the unavoidable dissipative system-environment interaction, which tends to asymptotically diminish entanglement. (c) $\mathcal{L}_{\text{max}}$ as a function of $p_e$ for an initial incoherent (yellow) and a coherent (purple) spin. The difference in entanglement (dashed curve) between the two cases attains the highest value 0.32 when $p_e=0.43$. These three curves are independent of the choice of $m$ and $g$. For all the panels, $\Delta = 0$.}
    \label{fig_LN_Osc_vac}
\end{figure*}

In the former case, obtaining a closed-form expression for $\mathcal{L}$ even in this simpler case is somewhat intricate, as the eigenvalues of the partial transpose of $\rho_{\text{b}}^{\text{sup}}(t)$ lack a simple algebraic structure. However, we have numerically verified that out of the four eigenvalues, only one becomes negative, contributing to $\mathcal{L}$. In the latter case, however, we were able to derive a closed-form expression for $\mathcal{L}$ (see Appendix~\ref{sec:log_neg} for details), and it is given by
\begin{align}
   \mathcal{L} = &\log_2\Big[p_e\sin^2\tau  \nonumber \\
    &\quad\quad+\sqrt{\left(1-p_e\sin^2\tau\right)^2 + g_{12}^2 p_e^2\sin^4\tau}\Big],
\label{eqn_LN_th_qbt_vac_osc}
\end{align}
where $\tau=\sqrt{m! (g_1^2+g_2^2)} t$, and $g_{12} = 2g_1 g_2/(g_1^2+g_2^2)$.

In Fig.~\ref{fig_LN_Osc_vac}(a) we have displayed the variation of the entanglement between the two bosonic modes $\mathcal{L}$ for different system parameters for the initial coherent as well as noisy incoherent spin. As mentioned before, we set $\Delta = 0$ for simplicity (see, however, Appendix~\ref{sec:coupled_eqns} where the role of detuning is analyzed in detail). Bearing in mind that $\mathcal{L}$ is upper-bounded by unity for these two initial cases (since there is only one initial excitation in the total system and both bosonic modes can be effectively treated as two-level systems), some interesting observations are in order: 

First, independent of the initial spin energy, the entanglement is genuinely non-Gaussian. That is, the entanglement contained in corresponding reference two-mode Gaussian states $\rho_{\text{Gauss}}(t)$, denoted by $\mathcal{L}_{\text{Gauss}}$, is identically zero at all times for both cases.
Using Simon's criterion~\cite{Simon_PRL_2000}, in Appendix~\ref{sec:log_neg_gauss}, we have analytically proved the separability of the reference two-mode Gaussian states $\rho^{\text{sup}}_{\text{Gauss}}(t)$ and $\rho^{\text{th}}_{\text{Gauss}}(t)$, both of which are constructed from the time-dependent covariance matrices of $\rho_{\text{b}}^{\text{sup}}(t)$ and $\rho_{\text{b}}^{\text{th}}(t)$, respectively.

Second, for both the initial states, the temporal entanglement evolution is completely periodic and has the same time period for similar values of $m$ and $g = g_1 = g_2$. The perfect oscillatory dynamics continues for any higher values of $m$with the oscillation frequencies scaled by a factor of $\sqrt{m!}$, while, for the same values of $p_e$, the amplitude of oscillation or the maximum achievable entanglement remains the same.
This clearly shows that there is no enhancement in the amount of entanglement possible by simply considering a higher-order JC interaction for these two initial cases. Nonetheless, the benefit lies in the fact that maximal entanglement can be achieved at a much shorter interaction time (to be precise, an improvement of $t/\sqrt{m!}$) and with greater frequency as the parameter $m$ increases to higher values. This attribute becomes significantly advantageous when accounting for the unavoidable dissipative interaction between the system and its environment, which tends to asymptotically diminish entanglement.
Similarly, for fixed $m$, the frequency of oscillations increases with increase in $\tilde{g} = \sqrt{g_1^2+g_2^2}$ and once again, if $p_e$ remains the same, the extent of the entanglement is the same. Therefore, similar reasoning holds even for going from a weak to a strong coupling regime. Interestingly, whenever $\mathcal{L}$ becomes zero, the global state returns to its initial separable state $\ket{e,\, 0,\, 0}$.  

For both initial cases, independent of the chosen value of $m$, $p_e$ controls the maximum achievable entanglement, as given by $\mathcal{L}_{\text{max}}^{\text{sup}} =\log_2[1 + p_e]$, and $\mathcal{L}_{\text{max}}^\text{th} =\log_2\big[p_e + \sqrt{1-2p_e +  2p_e^2}\big]$, respectively (see Appendix~\ref{sec:log_neg}). Maximum entanglement ($\mathcal{L}_{\text{max}}=1$) is reached when the qubit is in the excited state, i.e., $p_e = 1$ or $\phi=\frac{\pi}{2}$. The dependence of $\mathcal{L}_{\text{max}}$ on $p_e$ can more clearly be seen in Fig.~\ref{fig_LN_Osc_vac}(c) for both initial spin states. The advantage of having a coherent resource over an incoherent one for the production of entanglement (dashed curve) counterintuitively attains the highest value of 0.32 when $p_e=0.43$ (as opposed to $p_e=0.5$).

Another interesting aspect of the dynamics pertains to the periodic ``death'' of entanglement over a relatively longer period before it rises each time. Intuitively, this is related to the fact that the oscillators do not interact directly; the entanglement between them is always mediated via the qubit. Naturally, it takes a while for the correlations between the oscillators to build up. Mathematically, this can be explained by examining the short-time limit of $\mathcal{L}$ in Eq.\eqref{eqn_LN_th_qbt_vac_osc}, which reveals that $\mathcal{L}$ exhibits a~$\tau^4$ type nonlinear scaling during the initial phase of the evolution. To gain further insights, we compared the rise in the oscillator-oscillator entanglement to that of the qubit-oscillator entanglement. We found that for the latter, there is a linear rise during the initial phase. The details of this analysis can be found in Appendix~\ref{sec:log_neg}.

Finally, we note that $g_1=g_2$ is the optimal choice to get maximum entanglement.  This is due to the symmetry of the tripartite quantum system under the exchange of the bosonic modes $a_1 \leftrightarrow a_2$ (see Appendix~\ref{sec:log_neg} for further details). 

Up to this point, we have quantified the extent and non-Gaussian nature of entanglement between the two bosonic modes without thoroughly exploring the specific nature of the obtained states. Specifically, can distinctive two-mode entangled states be generated from $\rho_{\text{b}}^{\text{sup}}(t)$ and $\rho_{\text{b}}^{\text{th}}(t)$? In the following, we demonstrate, as an application, that with appropriately selected system parameters and initial state preparation, we can generate any desired target NOON states, expressed as follows
\begin{equation}
    \ket{\psi}_{\text{f}} = \frac{1}{\sqrt{2}}\left(\ket{N\,0} + \ket{0\,N}\right).
    \label{eqn_noon}
\end{equation}

To demonstrate this, we compute the fidelities, denoted as $\mathcal{F}$, of the states $\rho_{\text{b}}^{\text{sup}}(t)$ and $\rho_{\text{b}}^{\text{th}}(t)$ with the target state $ \ket{\psi}_{\text{f}}$. The fidelity is calculated using the expression $\mathcal{F} = {}_{\text{f}}\bra{\psi} \rho \ket{\psi}_{\text{f}}$~\cite{nielsen_chuang_2010}. Interestingly, it is observed that for both entangled states, the fidelities are identical, as given by
\begin{align}
    \mathcal{F} = \frac{1}{2} \left(|x_4|^2 + x_3x_4^* + x_3^*x_4 + |x_3|^2\right).
\end{align}
If the spin is initially prepared in the excited state and we set $g_1=g_2$, it becomes evident that at intervals of $t = \pi/(2\sqrt{m!}\tilde{g})$, the desired NOON state is generated. Notably, with a judiciously chosen order of the nonlinear JC interaction $m$, any specific target NOON state can be engineered {\em deterministically}. The other way to produce arbitrary NOON states would be through a measurement on the spin which is {\em probabilistic}. For the initial coherent spin state $|\psi(t) \rangle^{\rm sup}$, measuring the ground state of the spin results in
\begin{align}
\langle g\ket{\psi(t)}^{\text{sup}} = x_1\ket{0, 0} + x_3 \ket{m, 0} + x_4 \ket{0, m}.    
\end{align}
On the other hand, a similar measurement for the initial noisy spin would give
\begin{align}
\langle g|\rho^{\rm th}(t)|g \rangle &= |x_1|^2 |0,0 \rangle \langle 0,0| + |x_3|^2 |m,0 \rangle \langle m,0| \nonumber \\
&\,\,\,+ |x_4|^2 |0,m \rangle \langle 0,m| + x_3x_4^* |m,0 \rangle \langle 0,m| \nonumber \\
&\,\,\,+ x_3^* x_4|0,m \rangle \langle m,0|.
\end{align}
Evidently, if we choose $\phi=\pi/2$, we end up with the desired NOON state. It is worth mentioning at this point the existence of protocols for the generation of NOON states through multiphoton interactions of different kinds in ion traps~\cite{Zou_PRA_2001} as well as in systems with two cavity fields interacting with a superconducting qubit~\cite{Strauch2010,Zhao_sci_rep_2016}. We mention in passing that it is also possible to engineer a genuine tripartite entangled W state by exploiting the system parameters. The details of the analysis can be found in Appendix~\ref{sec:w_state}.\\

\noindent{\bf Spin coherence:} To understand the dynamics of spin coherence $\mathcal{C}(t)$, we first deduce the reduced density matrices for the two-level system by tracing over the bosonic modes. For the initial superposition spin state, we have
\begin{align}
  \rho_{\text{s}}^{\text{sup}}(t) = 
    \begin{pmatrix}
        \vert x_1\vert^2 +\vert x_3\vert^2 +\vert x_4\vert^2 & x_1 x_2^* \\
        x_1^* x_2 & \vert x_2\vert^2 
    \end{pmatrix}.
    \label{eqn_cq_sup_vac_osc}
\end{align}
On the other hand, for the initial thermal spin state, we get
\begin{align}
  \rho_{\text{s}}^{\text{th}}(t) = 
    \begin{pmatrix}
        \vert x_1\vert^2 +\vert x_3\vert^2 +\vert x_4\vert^2 & 0 \\
        0 & \vert x_2\vert^2 
    \end{pmatrix}.
    \label{eqn_cq_th_vac_osc}
\end{align}

 In the former case, $\mathcal{C}$ displays oscillatory dynamics (akin to the behavior observed in bosonic entanglement $\mathcal{L}$) with the maximum achievable coherence $\mathcal{C}_{{\text{max}}}$ being limited by the initial coherence of the spin, as can be easily seen analytically from Eq.\eqref{eqn_cq_sup_vac_osc}. The initial Bloch vector, in this case, performs nontrivial rotations on the Bloch sphere, i.e.,
\begin{equation}
  \begin{pmatrix}
    \sin(2\phi) \\
    0 \\
    \cos(2\phi)
  \end{pmatrix}^\top
      \longrightarrow  
  \begin{pmatrix}
    \sin(2\phi)\cos(\sqrt{m!}\tilde{g}t) \cos(\omega_0 t) \\
    -\sin(2\phi)\cos(\sqrt{m!}\tilde{g}t) \sin(\omega_0 t) \\
    1-2\sin^2\phi \, \cos^2(\sqrt{m!}\tilde{g}t)
  \end{pmatrix}^\top\,.
 \end{equation} 
In complete contrast to the bosonic entanglement, if the two-level system is initially prepared incoherently (thermal noisy spin), the Hamiltonian fails to induce any subsequent coherence in the spin. This observation is evident from Eq.\eqref{eqn_cq_th_vac_osc}, as the coherence terms are simply zero. Consequently, the Bloch vector undergoes a trivial rotation around the $\sigma_z$ axis on the Bloch sphere given by $\left(0,\, 0,\, 1-2p_e\right)\longrightarrow \left(0,\,0,\,1-2p_e \, \cos^2(\sqrt{m!}\tilde{g}t)\right)$. 
Thus, the insights derived from the $m$-photon JCM further extend and contextualize the findings presented in~\cite{laha_rabi_2023}, placing the present work within a broader and more generic framework.

\subsubsection{\label{subsubsec:3} `Dissipative-like' effects of additional Kerr nonlinearities}
In our investigation of the \textcolor{blue}{$m$}-photon JCM, we have discerned that the emergence of bosonic entanglement is contingent upon the initial state of the spin when both bosonic modes are prepared in their ground states. A comparative analysis has revealed that, regardless of the chosen value for $m$, achieving maximum entanglement ($\mathcal{L}_{\text{max}}=1$) is feasible only when the two-level system is initialized in the excited state. Additionally, we found that as the full system evolves under such a Hamiltonian, the spin remains incoherent if it is prepared incoherently.

Next, we consider the effect of additional Kerr nonlinearities on the system dynamics. It is worthwhile to mention that Kerr nonlinearity plays a significant role in generating nonclassical states~\cite{Shchesnovich_PRA_2011}, implementing quantum gates~\cite{Chono_PRR_2022}, exploring nonlinear optical processes~\cite{Bertet_2012}, quantum optomechanics~\cite{Lu_sci_rep_2013}, quantum error correction~\cite{Darmawan_PRXQ_2021,Kwon_npjQI_2022}, and quantum simulation and computation~\cite{Combes_PRA_2018}. In the present context, the inclusion of such Kerr nonlinearity in other multiphoton JCMs has been shown to play an important role in controlling the dynamics~\cite{Baghshahi_Laser_phys_2014,Ouyang_Chin_phys_2010,Liu_ijtp_2018,Naim_jrlr_2019,Singh_laser_phys_2019,singh_ijtp_2019}. 
The MPJC Hamiltonian including the Kerr nonlinearities in both the bosonic modes is given by 
\begin{align}
    H_{\text{Kerr}} = H + \sum_{i=1}^2 \chi_i \,a_i^{\dagger\,2} a_i^2,
    \label{mpjcm_kerr}
\end{align}
where $\chi_i$ is the strength of the nonlinearity of mode~$i$.

Even with the additional Kerr nonlinearities, the state vector of the tripartite system at any later time $t$ for initial superposition spin state and ground state oscillators can be described as in Eq.\eqref{eqn_psi_full_sup_qbt_vac_osc} but with modified time-dependent coefficients. The derivation of closed-form solutions for these coefficients for the most generic case proves intricate, as elucidated in Appendix~\ref{sec:kerr_eqns}. Consequently, numerical methods are employed to gain insights. However, for the special symmetric case, that is, when $\chi_1=\chi_2$, we could obtain exact analytical expressions for the time-dependent coefficients and are presented in Eq.\eqref{eqn_x1_x4_delta} with the identification $\Delta=-\chi(m^2-m)$. A parallel approach is adopted for the scenario involving an initial thermal spin qubit. We maintain $g_1=g_2=\frac{1}{\sqrt{2}}$ and $\Delta=0$ for consistency.\\

\noindent{\bf Bosonic entanglement:}
Analyzing the coupled differential equations in Appendix~\ref{sec:kerr_eqns}, it is clear that for $m=1$, the dynamics of $\mathcal{L}$ and $\mathcal{C}$ would remain unchanged as if the system does not feel the presence of the Kerr nonlinearities at all. This holds true for both single-sided and both-sided Kerr nonlinearities. This characteristic has been effectively leveraged in photonic quantum information processing, enabling the engineering of a controlled sign shift gate (CZ) on two single-rail qubits~\cite{peter_Laser_photonics_2011}. However, for higher $m$, the dynamics becomes sensitive to the presence of nonlinearities. 
\begin{figure}[ht!]
    \centering    \includegraphics{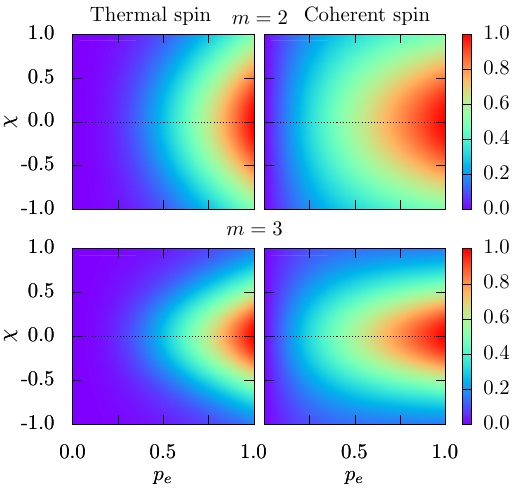}
    \vspace{-4ex}
    \caption{For an additional both-sided uniform Kerr nonlinearities, $\chi_1 = \chi_2 = \chi$ in the two-mode $m$-photon JCM (see $H_\text{Kerr}$ in Eq.\eqref{mpjcm_kerr}), the maximum bosonic entanglement $\mathcal{L}_{\text{max}}$ is shown as functions of $p_e$ and $\chi$, for $m=2$ (top row) and 3 (bottom row).  Both the bosonic modes are initially prepared in their respective ground states. The left (right) column corresponds to the initial noisy (coherent) spin state. Interestingly, the time evolution of the bosonic entanglement continues to be perfectly oscillatory as before.  But both the frequency and the amplitude of oscillations change as we gradually increase the magnitude of $\chi$. Here, $\mathcal{L}_{\text{max}}$ denotes the amplitude of oscillations. For fixed $p_e$, an increase in the magnitude of $\chi$ decreases the amplitude of oscillations $\mathcal{L}_{\text{max}}$ in a symmetric manner, regardless of the initial qubit state. In this sense, the added Kerr nonlinearity acts as a `dissipative-like' element in the entanglement production, even though the full system evolves unitarily. As expected, even in the presence of the Kerr nonlinearities, the initial coherent spin outperforms the noisy initial spin preparation. Also, for fixed $p_e$ and $\chi$, increase in $m$ decreases $\mathcal{L}_{\text{max}}$. Interestingly, for $m=1$, the entanglement dynamics remain unaffected as can be evidenced by analytics. Across all panels, uniform values of $g_1$, $g_2$ (set to $\frac{1}{\sqrt{2}}$), and $\Delta$ (set to 0) are maintained. The dashed lines in each panel simply correspond to $\chi=0$.}
    \label{fig_LNmax_Osc_vac_kerr}
\end{figure}

\begin{figure}[ht!]
    \centering
    \includegraphics{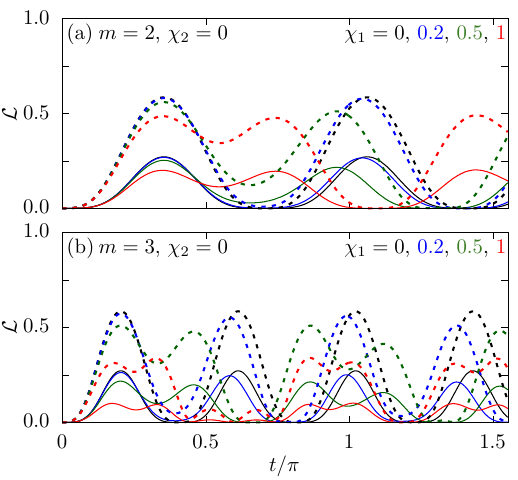}
    \vspace{-4ex}
    \caption{The role of additional single-sided Kerr nonlinearity in the two-mode $m$-photon JCM ($H_\text{Kerr}$ in Eq.\eqref{mpjcm_kerr}) on the bosonic entanglement $\mathcal{L}$ is shown for (a) $m=2$ and (b) 3, respectively. As before, the solid (dashed) lines correspond to the initial noisy thermal (coherent) spin state with $p_e = \sin^2\phi = 0.5$, and the bosonic modes are initially prepared in their respective ground states. Different colors correspond to different $\chi_1$ values while we set $\chi_2=0$. Also, $g_1=g_2=\frac{1}{\sqrt{2}}$ and $\Delta=0$. Due to the lack of symmetry in $H_\text{Kerr}$, the perfect oscillatory dynamics diminishes gradually as $\chi_1$ is increased. Similar to the both-sided case, here too, the added Kerr nonlinearity acts as a `dissipative-like' agent in terms of achieving higher entanglement.}
    \label{fig_LN_Osc_vac_1Skerr}
\end{figure}

For both-sided uniform Kerr nonlinearities, i.e., when $\chi_1 = \chi_2 = \chi$, we have verified that although the qualitative entanglement dynamics remains similar, i.e., perfect oscillations continue, both the frequency and the amplitude of oscillations of $\mathcal{L}$ change as we gradually increase the magnitude of $\chi$. In Fig.~\ref{fig_LNmax_Osc_vac_kerr}, we have analyzed the magnitude of these oscillations (denoted by $\mathcal{L}_{\text{max}}$) as functions of both $p_e$ and $\chi$. Independent of both the choice of initial spin state (noisy or coherent) and the sign of the Kerr nonlinearity $\chi$, for fixed $p_e$, departure from $\chi=0$ decreases $\mathcal{L}_{\text{max}}$.  Hence, the introduced Kerr nonlinearities function as a `dissipative-like' element in the entanglement production process, despite the complete dynamics of the system remaining unitary. The exact analytical solution indicates that the added nonlinearity takes the system out of resonance and consequently results in the reduction of the entanglement. As expected, even in the presence of the Kerr nonlinearities, the coherent qubit outperforms the thermal qubit. Also, for fixed $p_e$ and $\chi$, increase in $m$ decreases $\mathcal{L}_{\text{max}}$.

 The symmetry in the Hamiltonian is broken once we consider one-sided Kerr nonlinearity by setting either $\chi_1=0$ or $\chi_2=0$. As a result, the oscillatory dynamics ceases to be perfect. In other words, the entanglement dynamics becomes increasingly complex with an increase in the strength of the nonlinearity for the single-sided case, as can be seen in Fig.~\ref{fig_LN_Osc_vac_1Skerr}. \\

\noindent{\bf Spin coherence:} 
In the case of an initially coherent spin, anticipated oscillatory dynamics is observed for additional both-sided Kerr nonlinearity. However, for single-sided cubic nonlinearity, the perfect oscillations gradually diminish with increasing nonlinearity strength. Importantly, even with the inclusion of Kerr nonlinearities, the two-level system remains incoherent when prepared incoherently. This aligns with expectations, as the qubit density matrix maintains a structure akin to Eq.\eqref{eqn_cq_th_vac_osc} with adjusted time-dependent coefficients. In the following, we explore the emergence of spin coherence even when initially prepared incoherently, leveraging additional dispersive coupling.

\subsubsection{\label{subsubsec:4}Emergence of spin coherence through additional dispersive couplings}
Thus far we have found that an incoherent spin alone cannot induce spin coherence if the quantum system is described by the $m$-photon JCM including additional nonlinear Kerr interactions. Now, in the context of a single-mode JCM, earlier work~\cite{laha_rabi_2023} has already demonstrated the insufficiency of thermal energy alone in instigating spin coherence and shown that it is imperative to introduce an additional dispersive spin-boson interaction, denoted by $g_{z}\sigma_{z} X$, where $X = (a+a^\dagger)/\sqrt{2}$ is the position quadrature of the bosonic mode and $g_{z}$ is the strength of the additional dispersive coupling. 
In spin-boson systems, for instance, such dispersive couplings, in addition to the $m$-photon JC interaction, have already been analyzed theoretically~\cite{Puebla_symm_2019} for a single bosonic mode. There, it has been shown that the dispersive coupling effectively shifts the spin frequency depending on the state of the corresponding bosonic mode. On the other hand, in superconducting circuits similar dispersive couplings are already demonstrated experimentally~\cite{Chiorescu_nature_2004,yoshihara_superconducting_2017}. In the following, we adopt a strategy similar to the one in~\cite{laha_rabi_2023} and broaden the scope of our findings to encompass a considerably more generalized model. 

The two-mode $m$-photon JC Hamiltonian including additional dispersive couplings is given by
\begin{align}
    H_{\text{disp}} &=  H + \sum_{i=1}^{2}g_{z_i}\, \sigma_{z} X_i,
    \label{eqn:ham_disp}
\end{align}
where $g_{z_i}$ ($i=1,\,2$) are the dispersive coupling strengths, and $X_i=(a_i+a_i^\dagger)/\sqrt{2}$.
For simplicity, we continue to assume $g_1 = g_2 = \frac{1}{\sqrt{2}}$ and $\Delta=0$. Obtaining an analytical solution for the state of the tripartite system at a later time $t$, starting from the initial state described in Eq.\eqref{eqn_rho0_th} and governed by the Hamiltonian in Eq.\eqref{eqn:ham_disp}, presents a complex challenge. Therefore, we turn to numerical simulations to draw meaningful inferences.

\begin{figure}[ht!]
    \centering
    \includegraphics{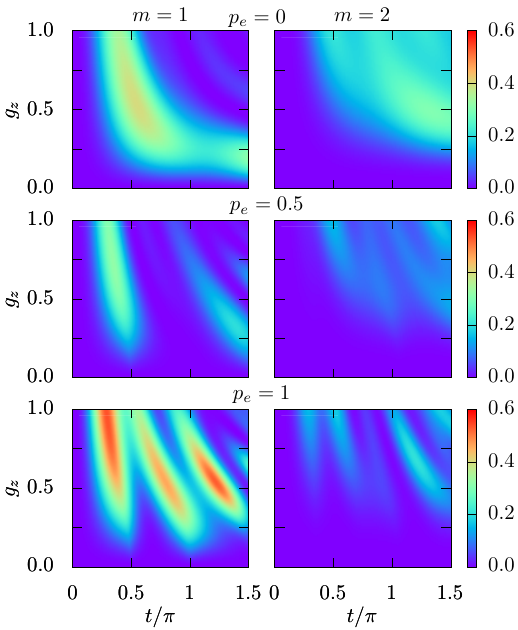}
    \vspace{-4ex}
    \caption{Visualization of spin coherence $\mathcal{C}$ arising exclusively from initially noisy and incoherent spins (with $p_e=0$, 0.5, and 1, corresponding to the top, center and the bottom rows, respectively) within the two-mode $m$-photon Jaynes-Cummings model, augmented by an additional dispersive spin-boson coupling (refer to Eq.\eqref{eqn:ham_disp}). The dependence on time and the dispersive coupling parameter $g_{z} = g_{z_1} = g_{z_2}$ is illustrated. Assuming the initial ground states for both bosonic modes, and setting $g_1 = g_2 = \frac{1}{\sqrt{2}}$, and $\Delta=0$ for consistency, the left and right columns represent cases for $m=1$ and 2, respectively. Notably, coherence emerges even when $p_e=0$ in the presence of additional dispersive couplings. However, an increase in the nonlinearity parameter $m$ does not necessarily enhance coherence production during the transient phase of the dynamics.}
    \label{fig_cq_Osc_vac_szx}
\end{figure}

Indeed, the inclusion of such an additional interaction remarkably generates coherence in the two-level system from initial thermal noise, as can be seen in Fig.~\ref{fig_cq_Osc_vac_szx}. We have verified that such an emergence in spin coherence is also visible even if $g_{z_1}=0$ or $g_{z_2}=0$. Similar to the findings reported in~\cite{laha_rabi_2023}, our results also reveal that the supplementary dispersive interaction, which individually produces only a trivial rotation of the Bloch vector around the $\sigma_z$ axis--much like the MPJC interactions--exhibits a nontrivial rotation of the Bloch vector when acting concurrently, thus inducing coherence in the spin. The temporal evolution of $\mathcal{C}$, however, is much more complex rather than perfectly oscillatory and also highly sensitive to the choice of $m$ and $p_e$, as shown in Fig.~\ref{fig_cq_Osc_vac_szx}.  Importantly, coherence is generated even when $p_e=0$. As expected, when the initial spin is in the excited state ($p_e=1$) the coherence production is maximum, while a maximally noisy spin ($p_e=0.5$) generates the least coherence.  

Contrary to expectations, we report that increasing the order of the nonlinearity parameter $m$ in MPJC coupling does not necessarily enhance the coherence production. We note that these observations are made based on the transient phase of the dynamics, although at later times, it may be possible that higher $m$ produces larger coherence. However, we should be careful as we are only analyzing here the dissipationless evolution, and when we consider the effects of the inevitable couplings to the environment (see Section~\ref{sec:results_open}) those effects will be greatly inhibited. Ultimately, as depicted in Fig.~\ref{fig_cq_Osc_vac_szx}, it is evident that the level of coherence generated tends to rise with the augmentation of $g_z$.

To provide a comprehensive overview, it is essential to note that the development of spin coherence is not contingent upon the presence of both bosonic modes; in fact, even a single bosonic mode is sufficient to induce spin coherence. The specifics of this investigation are discussed in Appendix~\ref{sec:single_mode}.

\begin{figure*}[ht!]
    \centering
    \includegraphics{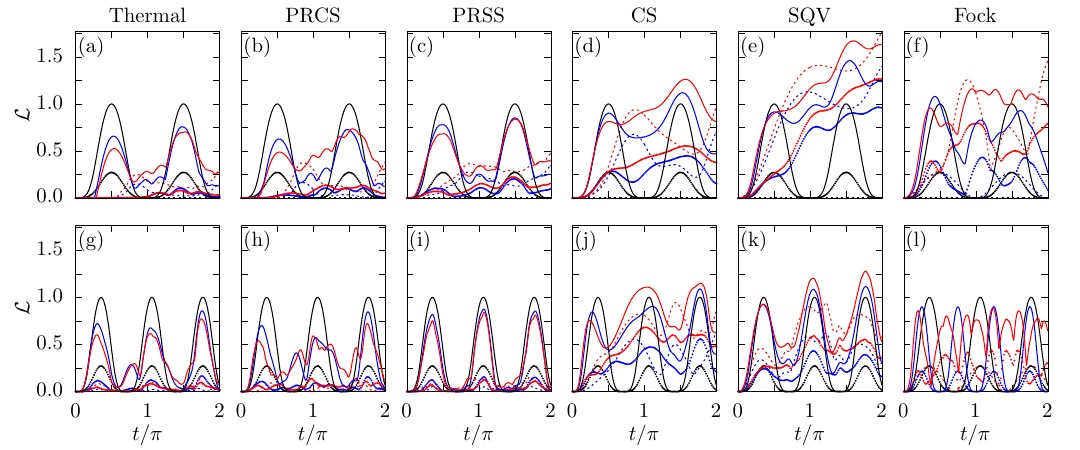}
    \vspace{-4ex}
    \caption{The temporal evolution of the bosonic entanglement $\mathcal{L}$ for the two-mode $m$-photon JCM governed by the Hamiltonian $H$ in Eq.\eqref{eqn_mpjcm} is displayed for various initial incoherent and coherent bosonic states. From left to right, these are the thermal state, the phase-randomized coherent state (PRCS), the phase-randomized squeezed state (PRSS), the standard coherent state (CS), the squeezed vacuum state (SQV), and the Fock state with $\bar{n}=$ 0 (black), 1 (blue), and 2 (red), respectively. The spin is considered to be initially in a noisy thermal state with $p_e=0$ (dashed), 0.5 (dotted), and 1 (solid curves), respectively. We continue to assume that the second bosonic state is still prepared in the ground state. The top (bottom) row corresponds to $m=1$ ($m=2$). We have numerically verified that the entanglement in all these cases continues to be genuinely non-Gaussian. For consistency, we work with $g_1 = g_2 = \frac{1}{\sqrt{2}}$, and $\Delta=0$.}
    \label{fig_LN_nbar_qbt_vac}
\end{figure*}

\subsection{\label{subsec:2}Coherent versus incoherent initial bosonic states}
In our analysis so far, we have assumed that both bosonic modes are initially prepared in ground states before the temporal evolution begins. In the following, we will relax this assumption and explore the changes that arise in the dynamics, when the bosonic states are prepared otherwise. To simplify the analysis, we initially assume that one of the bosonic modes (say, the second one) is still prepared in the ground state $\ket{0}$. Further, the two-level system is assumed to be initially prepared in the thermal state only. Also, we consider the clean two-mode $m$-photon JCM, i.e., $H$ in Eq.\eqref{eqn_mpjcm} for the analysis. 

Now, concerning the first bosonic mode, various coherent and incoherent initial states can be considered such as the standard coherent state (CS) $\ket{\alpha}$, the squeezed vacuum state $\ket{\xi}$, and incoherent ones such as the thermal state $\rho_{\text{th}}$ with mean thermal energy $\bar{n}$. For simplicity, we assume both $\alpha$ and $\xi$ to be real.  
For completeness, given below are the Fock basis $\{\ket{n}\}$ expansions of these states
\begin{align}
 \displaystyle \ket{\alpha} &= e^{-\frac{1}{2}\vert\alpha\vert^2}\sum_{n=0}^{\infty} \frac{\alpha^{n}}{\sqrt{n!}}  \ket{n},\\
 \displaystyle \ket{\xi} &= \frac{1}{\sqrt{\cosh{r}}} \sum_{n=0}^{\infty} (-\tanh{r})^n e^{-i n\theta}\frac{(2n)!}{2^n n!}  \ket{2n},\\
 \displaystyle \rho_{\text{th}} &= \frac{1}{1+\bar{n}}\sum_{n=0}^{\infty} \left(\frac{\bar{n}}{1+\bar{n}}\right)^{n}  \ket{n}\bra{n}.
\end{align}
Apart from these, we have also considered the phase-randomized coherent state (PRCS), defined as
\begin{align}
 \displaystyle \rho_{\text{PRCS}} = \int_{0}^{2\pi} \frac{d\theta}{2\pi} \, \ket{\alpha e^{i\theta}}\bra{\alpha e^{i\theta}} = e^{-\vert\alpha\vert^{2}} \sum_{n=0}^{\infty} \frac{\vert\alpha\vert^{2n}}{n!}\ket{n}\bra{n}.
   \label{eqn_prcs}
\end{align}
Similarly, phase-randomized squeezed state (PRSS) can be obtained from the squeezed vacuum states and is given by
\begin{align}
   \displaystyle \rho_{\text{PRSS}} = \frac{1}{\cosh{r}} \sum_{n=0}^{\infty} (-\tanh{r})^{2n} \left(\frac{(2n)!)}{2^n n!}\right)^2  \ket{2n}\bra{2n}.
   \label{eqn_prss}
\end{align}
By definition, both PRCS and PRSS are therefore incoherent in the photon number basis. For completeness, we also consider the highly nonclassical excited Fock states $\ket{n}$.
For ease of comparison, we denote by $\bar{n}$ the mean energy of all the states, i.e., $\bar{n} = \vert\alpha\vert^2 = \sinh^2{r}$. It is interesting to note that all these CV states are reduced to the vacuum $\ket{0}$ in the limit $\bar{n}=0$.

Employing numerical analysis, the impact of utilizing various coherent and incoherent oscillator energies on the temporal evolution of bosonic entanglement $\mathcal{L}$ is depicted in Fig.~\ref{fig_LN_nbar_qbt_vac}. The black curves in all panels, representing $\bar{n}=0$, align precisely with the corresponding curves depicted in Fig.~\ref{fig_LN_Osc_vac}.
In addition to this obvious one, several key observations can be drawn from Fig.~\ref{fig_LN_nbar_qbt_vac} as discussed in the following: 

Even when $p_e=0$, the entanglement between the two bosonic modes persists across all scenarios, gradually strengthening with increasing $\bar{n}$, as depicted by the dashed curves. Notably, the onset of entanglement may be delayed and tends to be weaker for initial thermal and PRCS states relative to other scenarios, while the squeezed vacuum state $\ket{\xi}$ yields maximal entanglement. Contrary to intuition, for $p_e=0$, the augmentation of $m$ in the MPJC interaction results in lower values of $\mathcal{L}$ across all scenarios. Specifically, for higher Fock states $\ket{n}$, the entanglement is generated when $n\geqslant m+1$. However, the initiation of the entanglement occurs more rapidly as $m$ is increased to higher values for both the initial thermal and PRCS states.

As illustrated by the dotted curves in all panels, corresponding to a maximally mixed spin state (i.e., when $p_e=0.5$), an escalation in the noise within the initial spin leads to a diminished entanglement across all scenarios except for the higher Fock states. As opposed to $p_e=0$, now the two modes become entangled for all higher Fock states, irrespective of the relation between $n$ and $m$ mentioned above. Interestingly, for all initial mixed bosonic states, the production of entanglement is adversely affected by an increase in the mean energy $\bar{n}$ during the transient phase of dynamics. Conversely, the behavior is the opposite for initial pure bosonic states. Increasing $m$ in this case, serves a qualitatively similar role in the entanglement production, mirroring the effects observed when $p_e=0$. 

A significant amount of entanglement is generated even for initial thermal or PRCS states when $p_e=1$, as depicted by the solid curves. The influence of $\bar{n}$ and $m$ on the entanglement dynamics remains similar to the case $p_e=0.5$.

Independent of the values of $p_e$, $\mathcal{L}$ no longer exhibits perfect oscillations for all the different initial bosonic states, indicating the involvement of more energy levels beyond just $\ket{0}$ and $\ket{m}$ for each oscillator state when $\bar{n} \ne 0$. The only exception is the initial Fock state $\ket{n}$ with $n=m$ for which the dynamics remains perfectly oscillatory. Consequently, the upper bound of $\mathcal{L}$ is no longer constrained to unity. Notably, $\mathcal{L}$ surpasses unity for higher resourceful initial bosonic states (see last three columns in Fig.~\ref{fig_LN_nbar_qbt_vac}. However, when starting with incoherent oscillator states, we find that $\mathcal{L}$ remains less than or equal to 1, particularly during the transient phase of the dynamics.  

Lastly, and of greater significance, we have numerically confirmed that the entanglement remains genuinely non-Gaussian across all cases.

\begin{figure}[ht!]
    \centering
    \includegraphics{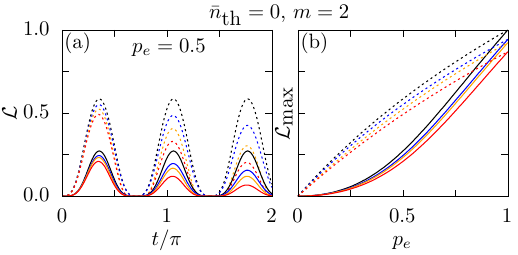}
    \vspace{-4ex}
    \caption{For the standard two-mode $m$-photon JCM, the impact of vacuum fluctuations (i.e., $\bar{n}_{\text{th}}=0$) on (a) the temporal evolution of bosonic entanglement $\mathcal{L}$ for $p_e = \sin^2\phi = 0.5$ and (b) $\mathcal{L}_{\text{max}}$ as a function of $p_e$,  corresponding to initial coherent (dashed curves) and incoherent thermal (solid curves) qubit states is depicted. For consistency, we use similar parameters and initial bosonic states as in Fig.~\ref{fig_LN_Osc_vac}. In both panels, the blue (yellow) curves represent the entanglement evolution in the presence of dissipation (dephasing) only, i.e., $\lambda_{d_b}=\lambda_{d_q}=0$ ($\lambda_{r_b}=\lambda_{r_q}=0$) and we set $\lambda_{r_b}=\lambda_{r_q}=0.05$ ($\lambda_{d_b}=\lambda_{d_q}=0.05$), while the red curves showcase the cumulative effects of both ($\lambda_{r_b}=\lambda_{r_q}=\lambda_{d_b}=\lambda_{d_q}=0.05$). The black curves correspond to unitary dynamics. Upon comparison, it is clear that, for similar decay rates, $\mathcal{L}$ is more adversely affected by dephasing than dissipation in both cases.}
    \label{fig_LN_open_sys_nth_0}
\end{figure}

\section{\label{sec:results_open}Open system dynamics}
Up until now, we have restricted ourselves to the perfect unitary evolution of the tripartite quantum system, thereby ignoring completely the crucial role of inevitable system-environment couplings on the system's dynamics. In the subsequent analysis, using numerical simulation, we present a comprehensive analysis of the degree to which the dynamics of bosonic entanglement and qubit coherence are influenced due to environmental-induced interactions.  We employ the standard Lindblad formalism that includes the Markovian approximation, among other assumptions.  In this framework, the evolution of the reduced density matrix of the tripartite system, denoted $\rho_{S}(t)$, is governed by the Lindblad master equation, given by
\begin{align}
 \displaystyle\frac{d\rho_{S}}{dt}  = -i \left[H, \rho_{S}\right] + \sum_{k} \left( L_{k} \rho_{S} L_{k}^{\dagger} -  \frac{1}{2}\left[\rho_{S}, L_{k}^{\dagger} L_{k}\right] \right).
 \label{rhot_decoh}
\end{align}
Here, $L_{k} = \sqrt{\lambda}_{k} A_{k}$ are the Lindblad operators, and the environment couples to the system through the operators $A_{k}$ with coupling rates $\lambda_{k}$. To keep the numerical analysis simpler, we assume that both the bosonic modes and the two-level system are coupled to the same thermal environment with temperature $\bar{n}_{\text{th}}$. Also, we note that in Eq.\eqref{rhot_decoh} we have ignored the Lamb shift contribution leading to a small renormalization of the system energy levels. 
Now, for the two bosonic modes, the Lindblad operators ($L_k$'s) are given by $\sqrt{\lambda_{r_i}(1+\bar{n}_{\text{th}})} \,a_i$, $\sqrt{\lambda_{r_i}\, \bar{n}_{\text{th}}}\, a_i^{\dagger}$, and $\sqrt{\lambda_{d_i}} a_i^\dagger a_i$, respectively ($i=1,\,2$). Further, we assume $\lambda_{r_1} = \lambda_{r_2} = \lambda_{r_b}$ and $\lambda_{d_1} = \lambda_{d_2} = \lambda_{d_b}$, for simplicity. Similarly, for the two-level system, the Lindblad operators are $\sqrt{\lambda_{r}(1+\bar{n}_{\text{th}})} \,\sigma_-$,  $\sqrt{\lambda_{r_q}\, \bar{n}_{\text{th}}}\, \sigma_+$ and $\sqrt{\lambda_{d_q}}\sigma_+\sigma_-$, respectively. Note that while the dissipation and relaxation rates depend on the temperature ($\bar{n}_{\text{th}}$) of the bath, the dephasing rates are independent of $\bar{n}_{\text{th}}$.

\begin{figure}[ht!]
    \centering
    \includegraphics{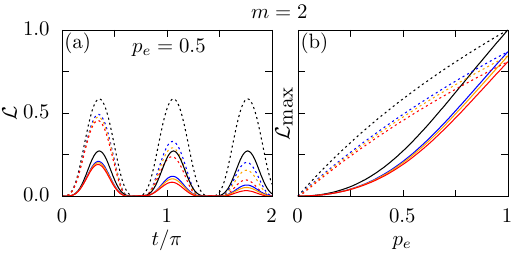}
    \vspace{-4ex}
    \caption{For the standard two-mode $m$-photon JCM, the effect of finite bath temperature $\bar{n}_{\text{th}}$ on (a) the temporal evolution of bosonic entanglement $\mathcal{L}$ for $p_e = 0.5$ and (b) $\mathcal{L}_{\text{max}}$ as a function of $p_e$, corresponding to initial coherent (dashed curves) and incoherent thermal (solid curves) qubit states is depicted. For consistency, we use similar parameters and initial bosonic states as in Fig.~\ref{fig_LN_Osc_vac}. In both panels, the blue, yellow, and red curves correspond to $\bar{n}_{\text{th}}=$ 0, 0.2, and 0.5, respectively. Further, $\lambda_{r_b}=\lambda_{r_q}=\lambda_{d_b}=\lambda_{d_q}=0.05$. The black curves correspond to unitary dynamics. }
    \label{fig_LN_open_sys_nth}
\end{figure}

We first assume $\bar{n}_{\text{th}}=0$ so that the quantum correlations suffer only due to vacuum fluctuations (loss). We set the system parameters similar to those in Fig.~\ref{fig_LN_Osc_vac}(a) to make a comparative statement. In addition to the expected continuous asymptotic decrease in the extent of entanglement due to vacuum fluctuations, our numerical analysis reveals that for similar decay rates, $\mathcal{L}$ is more adversely affected by dephasing than dissipation (see Fig.~\ref{fig_LN_open_sys_nth_0}). This observation holds true for both scenarios corresponding to the initial superposition and thermal qubit states. These conclusions are further augmented by the behavior of $\mathcal{L}_{\text{max}}$ as a function of $p_e$ as shown in Fig.~\ref{fig_LN_Osc_vac}(b). Here, the amplitude of the first oscillation serves as our chosen measure for $\mathcal{L}_{\text{max}}$. 

Finally, in Fig.~\ref{fig_LN_open_sys_nth} we show the effect of finite bath temperature $\bar{n}_{\text{th}}$ on the dynamics. Now, all the Lindblad operators contribute to the system dynamics. We find that a gradual increase in $\bar{n}_{\text{th}}$ results in a further reduction of the entanglement. 
For completeness, we also mention that similar reductions in the amount of bosonic entanglement and qubit coherence are also observed in all other cases that are reported in this work.

\section{\label{sec:conc}Conclusions}
Resource-efficient generation and optimal distribution of bosonic entanglement and spin coherence are fundamental building blocks of contemporary quantum technology. In this work, we have considered a fundamental spin-boson system from which we obtained (after a series of approximations, as explained in Appendix~\ref{sec:derivation}) a tripartite quantum system involving only two bosonic modes interacting with a single two-level system via the nonlinear $m$-photon JC interaction.

In the initial part, assuming that the bosonic modes are initially in the ground states, we have analytically solved the system Hamiltonian for both initial noisy and coherent spin states. We have found that the entanglement between the bosonic modes can be readily generated using solely noisy incoherent resources, thus proving to be remarkably resource-efficient. To this end, a detailed comparative analysis of the role of coherent and incoherent energy in producing entanglement has revealed that the latter operates with an expected lower efficiency, a challenge addressable through various distillation protocols. Furthermore, we have demonstrated the strategic advantage of choosing higher-order JC interactions for faster entanglement production, aiming to alleviate losses incurred from environmentally induced interactions. In addition, we have analytically proved that irrespective of the initial spin states, the generated entanglement is genuinely non-Gaussian, free from any Gaussian contribution. As a potential application, we have shown how to engineer arbitrary bipartite entangled NOON states or a genuine tripartite entangled W state leveraging the inherent multiphoton nonlinearity by choosing the system parameters appropriately. In optical quantum information science, these states are important resources~\cite{Dur2000,Dowling_2008,Bergmann_PRA_2016} and the former state has been widely used in quantum metrology~\cite{Giovannetti_nat_photon_2011}, and quantum lithography~\cite{Boto_PRL_2000}. However, our findings have also revealed that the nonlinearities within the MPJC interaction fall short of inducing coherence in the spin when it is prepared incoherently.

Expanding our analysis to incorporate the influence of additional Kerr nonlinearity on dynamics, we have found that increasing nonlinearity does not necessarily enhance entanglement production. Instead, it introduces a `dissipative-like' element in the dynamics, despite the full system dynamics being unitary.  Building on previous work~\cite{laha_rabi_2023}, we have highlighted the need for additional dispersive spin-boson interactions to achieve spin coherence from noisy initial states. Surprisingly, our numerical results have indicated that considering higher-order JC interactions may not be advantageous for generating higher coherence, at least during the transient phase of the dynamics. In the latter part, we have also investigated the role of coherent versus noisy initial bosonic modes on the dynamics. Finally, we have numerically investigated the detrimental role of environmentally induced effects on these phenomena. The numerical results presented here are obtained with the help of the {\it QuTiP} library~\cite{qutip}.

Our study lays the groundwork for understanding quantum correlations arising from both coherent and incoherent resources within a particular nonlinear quantum system. We anticipate that our findings will inspire further investigations into similar phenomena across diverse physical systems. Extending the tripartite model to include more oscillators and qubits or substituting the qubit with a multi-level discrete quantum subsystem are natural next steps. However, it is essential to note that such generalizations exponentially increase the combinations of different bipartite subsystems.
In these extended scenarios, investigating multipartite entanglement becomes particularly intriguing, offering insights into the intricate interplay of quantum correlations in complex quantum systems. Furthermore, the multiphoton generalization of these models introduces other intriguing phenomena, such as entanglement sudden death~\cite{Y_na__2006, laha_josab_2023} and coherence sudden death~\cite{Kaifeng_PRA_2016}. Another aspect of this study involves analyzing the nonclassicalities of the bosonic modes resulting from such nonlinear Hamiltonian evolution. This investigation is conducted in detail, considering a pure tripartite initial state where the spin is in a superposition state of the two basis states, and the two bosonic modes are in arbitrary Fock states~\cite{laha_jcm_wigner_2024}.


\begin{acknowledgments}
PL thanks Radim Filip and Darren W. Moore for useful discussions leading up to the work. We acknowledge funding from the BMBF in Germany (QR.X, PhotonQ, QuKuK, QuaPhySI) and from the Deutsche Forschungsgemeinschaft (DFG, German Research Foundation) – Project-ID 429529648 – TRR 306 QuCoLiMa (“Quantum Cooperativity of Light and Matter”).
\end{acknowledgments}

\appendix


\begin{figure*}[htbp]
\centering
\includegraphics[scale=0.45]{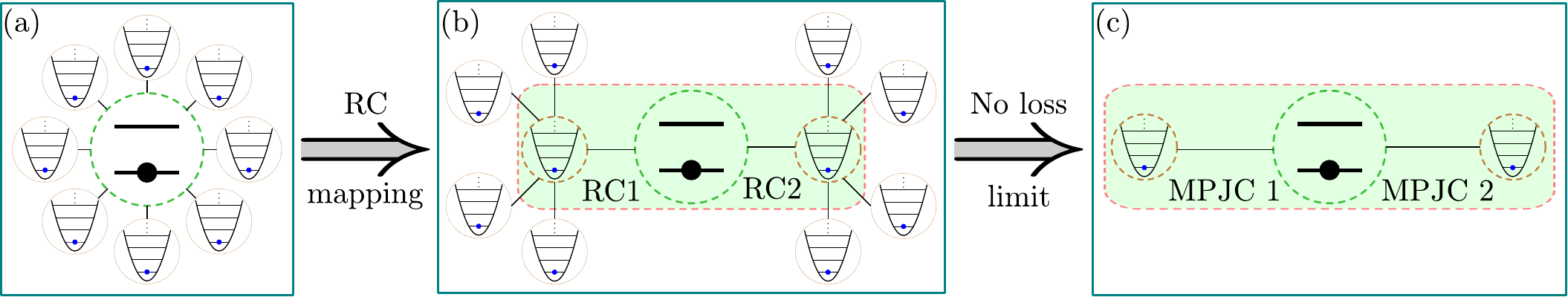}
\caption{(a) An illustrative representation of a generic spin-boson model, where a single spin (depicted as a two-level system) interacts with an environment composed of a large number of harmonic oscillators (bosonic modes), as described by the total Hamiltonian $H_\text{SB}$ in Eq.\eqref{eqn:H_SB}. One method to derive a two-mode MPJC Hamiltonian from $H_\text{SB}$ is through the utilization of the so-called `reaction-coordinate (RC) mapping', as elucidated in~\cite{Puebla_symm_2019}. This method entails a structured rearrangement of the environmental degrees of freedom to incorporate a small set of collective coordinates into the Hamiltonian, subsequently interacting with the remaining environment, as illustrated in panel (b). Finally, in the limit where dissipation effects are negligible enough to be disregarded, the resultant two-mode MPJC Hamiltonian $H$ in Eq.\eqref{eqn_mpjcm} is obtained.}
\label{Fig_flowchart}
\end{figure*}

\section{\label{sec:derivation} Realizing multiphoton Jaynes-Cummings interactions in spin-boson models}
In this Section, we briefly outline the key steps to realizing a two-mode $m$-photon JC Hamiltonian from the fundamental spin-boson interaction (details can be found in~\cite{Puebla_symm_2019}). Similar model Hamiltonian can also be realized in other quantum platforms such as the trapped ion~\cite{Leibfried_RMP_2003,Haffner_phys_rep_2008} and the superconducting circuit~\cite{Strauch2010,Felicetti_PRA_2018}. 

We begin with the conventional spin-boson Hamiltonian
\begin{align} 
 H_\text{SB} = H_\text{S} + H_\text{E} + H_\text{S-E},    
 \label{eqn:H_SB}
\end{align}
where $H_\text{S}= \frac{\omega'_0}{2} \sigma_z$ and $H_\text{E}= \sum_k \omega'_k c^\dagger_k c_k$, respectively denote the free-energy Hamiltonians associated with the two-level atom and the environment. The interaction between the system and the environment is characterized by $H_{\text{S-E}}=\sigma_x \sum_k f_k (c_k + c_k^\dagger)$. Here, $\omega'_0$ denotes the transition frequency between the two energy levels of the system, and $\sigma_i$ represents the Pauli operator with $i$ taking values $x$, $y$, or $z$. The frequency of the $k$th bosonic mode is denoted by $\omega'_k$, and $c_k$ and $c_k^\dagger$ stand for the annihilation and creation operators corresponding to this mode. The coupling strength between the system and the $k$-th bosonic mode is represented by $f_k$.

Now, employing the `reaction coordinate (RC) mapping'~\cite{thoss2001,smith2014} -- in which the relevant environmental bosonic modes are incorporated into the system to form an effective system Hamiltonian -- we can obtain a collective mode with associated annihilation and creation operators, $a$ and $a^\dagger$, such that
\begin{align} \label{ap3a}
\sum_k f_k (c_k + c_k^\dagger) = \lambda (a+a^\dagger).    
\end{align}
With this, we obtain a reaction coordinate Hamiltonian of the form\,\cite{smith2014,Puebla_symm_2019}
\begin{align} \label{ap3b}
H_\text{RC} &= \frac{\omega'_0}{2} \sigma_z + \lambda \sigma_x (a+a^\dagger) + \Omega a^\dagger a + \sum_k \omega_k b_k^\dagger b_k \nonumber \\ 
&\,\,\,+ (a+a^\dagger) \sum_k g_k (b_k+b_k^\dagger) + (a+a^\dagger)^2 \sum_k \frac{g_k^2}{\omega_k}, 
\end{align}
where $\lambda^2=\sum_k f_k^2$ and $\Omega=\lambda^{-1} \sqrt{\sum_k \nu_k f_k^2}$. The residual bath that contains creation\,($b_k^\dagger$) and annihilation\,($b_k$) operators is coupled to the reaction coordinate alone; it is characterized by an effective spectral density given by $J_\text{RC}(\omega) =\sum_k g_k^2 \delta(\omega-\omega_k)$. Starting with $H_\text{RC}$, the intermediate Hamiltonian takes the following form in the rotating frame: 
\begin{align} \label{ap3c}
H' = \omega a^\dagger a &+ \lambda \sigma_x (a+a^\dagger)   \nonumber \\
&+ \sum_{j=0}^{n_d} \frac{\epsilon_j}{2} [\cos (\Delta_j t) \sigma_z + \sin (\Delta_j t) \sigma_y],
\end{align}
with $\Delta_j$ being the detuning parameter. Performing another unitary transformation, that is, $H'' = \tilde{U} H' \tilde{U}^\dagger$, where $\tilde{U}$ is some combination of the displacement operator and the Pauli matrix, we obtain the Hamiltonian
\begin{align} \label{ap4a}
H'' = \omega a^\dagger a + \sum_{j=0}^{n_d} \frac{\epsilon_j}{2} \left(\sigma_+ e^{2\lambda (a-a^\dagger)/\omega} e^{-i\Delta_j t} + \text{H.C.}\right),    
\end{align}
where H.C. stands for Hermitian conjugate.
On making the rotating wave approximation in the Lamb-Dicke regime, we can end up with a Hamiltonian
\begin{align} \label{ap4b}
H_n = \frac{\tilde{\omega}}{2} \sigma_z &+ \tilde{\nu} a^\dagger a + \sum_{j \in R} \frac{\epsilon_j (2\lambda)^{n_j}}{2\Omega^{n_j} n_j!} \left(\sigma_+ a^{n_j} + \text{H.C.}\right) \nonumber \\
&\quad+ \sum_{j \in B} \frac{\epsilon_j (2\lambda)^{n_j}}{2\Omega^{n_j} n_j!} \left(\sigma_+ (-a)^{n_j} + \text{H.C.}\right). 
\end{align}
Here, the detuning parameters for the summations belonging to the sets $R$ and $B$ are $\Delta_{j\in R}=+n_j (\tilde{\nu}-\Omega) -\tilde{\omega}$ and $\Delta_{j\in B}=-n_j (\tilde{\nu}-\Omega) -\tilde{\omega}$, respectively. The desired multi-photon JCM term can be picked out with an appropriate choice of $\Delta$ in the $R$ set. Thus, we finally have the desired Hamiltonian
\begin{align} \label{ap5a}
H_1 = \frac{\omega_0}{2} \sigma_z + \omega a^\dagger a + g'_1 (\sigma_+ a^n + \sigma_- a^{\dagger\,n}).  
\end{align}
This Hamiltonian is precisely the multiphoton generalization of the Jaynes-Cummings model studied in Ref.~\cite{SUKUMAR_pla_1981}. Here, the atom makes a transition from the ground\,(excited) state to the excited\,(ground) state by absorbing\,(releasing) $n$-photons. 

Now, to derive a two-mode MPJC Hamiltonian, it is necessary to reorganize the initial environment in a way that preserves two collective coordinates within the expanded system. Each of these coordinates subsequently engages with its respective residual environment, as illustrated in Fig.~\ref{Fig_flowchart} (see~\cite{Puebla_symm_2019} for further details).



\section{\label{sec:log_neg} Details on Logarithmic Negativity}
\subsection{\label{sec:osc_osc_ent} Oscillator-oscillator entanglement}

For thermal qubit and ground state oscillators, the partial transpose of the reduced density matrix corresponding to the two bosonic modes $\rho_{\text{B12}}^{00}(t)$ in Eq.\eqref{eqn_rho_12_q_th} is  given by
\begin{align}
  [\rho_{\text{b}}^{\text{th}}(t)]^{\rm PT} = 
    \begin{pmatrix}
        \vert x_1\vert^2 + \vert x_2\vert^2 & 0 & 0 & x_3x_4^*\\
        0 & \vert x_4\vert^2 & 0 & 0\\
        0 & 0 & \vert x_3\vert^2 & 0 \\
        x_3^*x_4 & 0 & 0 & 0
    \end{pmatrix}.
\end{align}

\begin{figure}[ht!]
    \centering
    \includegraphics{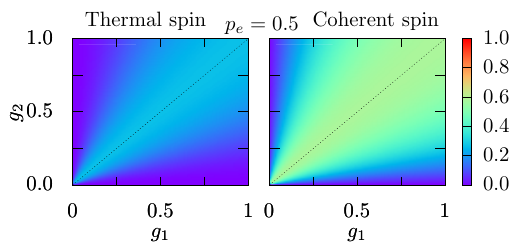}
    \vspace{-4ex}
    \caption{Illustration of the impact of coupling parameters $g_1$ and $g_2$ in Eq.\eqref{eqn_mpjcm} on the maximal achievable entanglement $\mathcal{L}_{\text{max}}$ (in other words, the amplitude of the perfect Rabi oscillations) is presented for both initial spin states. The parameters are set to $p_e=0.5$, $m=1$, and $\Delta=0$, with the bosonic modes initially prepared in their respective ground states. Notably, in both cases, optimal $\mathcal{L}_{\text{max}}$ is attained when $g_1=g_2$, as pointed out by the dashed lines.}
    \label{fig_g1g2_loop}
\end{figure}

The two trivial eigenvalues of $\rho^{PT}(t)$ are $|x_3|^2$ and $|x_4|^2$ which do not contribute to the logarithmic negativity. The other two eigenvalues are
\begin{align}
\lambda_{\pm} = \frac{1}{2}\left(\vert x_1\vert^2 + \vert x_2\vert^2 \pm \sqrt{(\vert x_1\vert^2 + \vert x_2\vert^2 )^2 + 4 \vert x_3\vert^2 \vert x_4\vert^2 }  \right).   
\end{align}
Evidently, $\lambda_{+}$ is always positive and $\lambda_{-}$ will be negative as long as $|x_3|^2 |x_4|^2 \neq 0$. Now, using the definition  $\mathcal{L} = \log_2\left[1+ 2 \left\vert\sum_{\lambda_i<0}\lambda_i \right\vert \right]$, we obtain
\begin{align}
 \mathcal{L} &= \log_2\left[1+ 2|\lambda_{-}|\right] = \log_2 \Big[ 1 - |x_1|^2-|x_2|^2 \nonumber \\
 & \quad\quad\quad\quad+  \sqrt{(\vert x_1\vert^2 + \vert x_2\vert^2 )^2 + 4 \vert x_3\vert^2 \vert x_4\vert^2 }  \Big].    
 \label{eqn_L_th_x1_4}
 \end{align}
Now, substituting $|x_1|^2+|x_2|^2 = 1-p_e\sin^2\tau$,  $|x_3|^2 = \frac{g_1^2}{g_1^2+g_2^2}p_e\sin^2\tau$, and $|x_4|^2 = \frac{g_2^2}{g_1^2+g_2^2}p_e\sin^2\tau$, where $\tau = \sqrt{m! (g_1^2+g_2^2)} t$, we get 
\begin{align}
\mathcal{L} =  \log_2\left[p_e\sin^2\tau + \sqrt{\left(1-p_e\sin^2\tau\right)^2 + g_{12}^2 p_e^2\sin^4\tau}\right]
 \label{eqn_ln_th}
\end{align}
where $g_{12}=\frac{2g_1 g_2}{g_1^2+g_2^2}$.
The parameter $g_{12}$ in Eq.\eqref{eqn_ln_th} is upper-bounded by unity, and the unit value is achieved when $g_1=g_2=g$, in which case
$\tau = \sqrt{2m!}gt$. The choice of equal coupling parameters to achieve the highest entanglement is further illustrated in Fig.~\ref{fig_g1g2_loop}. We find that this observation holds true for both initial spin states, regardless of the chosen order of the MPJC interaction parameter $m$.

Now, it is apparent from Eq.\eqref{eqn_ln_th} that selecting $\tau=(2n+1)\frac{\pi}{2}$, where $n=0,\,1,\,2\,\cdots$, yields the optimal time to get the highest entanglement. Substituting this into Eq.\eqref{eqn_ln_th} and setting $g_{12}=1$ we obtain the maximal achievable entanglement given by
\begin{align}
   \mathcal{L}_{\text{max}}^\text{th} &=\log_2\Big[p_e + \sqrt{\left(1-p_e\right)^2 +  p_e^2}\Big].
\end{align}
The equation above illustrates how the strength of the initial incoherent noise, denoted as $p_e$, influences the maximum attainable entanglement. It is evident that the maximum entanglement, $\mathcal{L}_{\text{max}}=1$, occurs when $p_e = 1$.

On the other hand, for the initial coherent spin state, we have
\begin{align} \label{c7}
[\rho_{\text{b}}^{\text{sup}}(t)]^{\rm PT} = 
\begin{pmatrix}
\vert x_1\vert^2 +\vert x_2\vert^2 & x_1 x_4^* & x_1^* x_3 & x_3x_4^* \\
x_1^* x_4 & \vert x_4\vert^2 & 0 & 0 \\
x_1 x_3^* & 0 & \vert x_3\vert^2 & 0 \\
x_3^*x_4 & 0 & 0 & 0
\end{pmatrix}.     
\end{align}
It can be verified that the eigenvalues are solutions of the characteristic equation
\begin{align}
&\lambda^4 - \lambda^3 + |x_2|^2 (|x_3|^2+|x_4|^2) \lambda^2 + [|x_3|^2 |x_4|^2 \nonumber \\
&\,\,\,\times (1-2|x_2|^2)] \lambda - |x_3|^4 |x_4|^4 = 0.    
\end{align}

As mentioned in the main text, the eigenvalues of  $[\rho_{\text{b}}^{\text{sup}}(t)]^{\rm PT}$ lack a simple algebraic structure. However, we have numerically verified that out of the four eigenvalues, only one becomes negative, contributing to $\mathcal{L}$. Nonetheless, using {\it Mathematica},  we could obtain the analytical expression for the maximal achievable entanglement in terms of the initial energy of the spin (setting $g_1=g_2=\frac{1}{\sqrt{2}}$ and $\tau = (2n+1)\frac{\pi}{2}$, similar to the earlier case), and is given by
\begin{align}
  \mathcal{L}_{\text{max}}^{\text{sup}} &=\log_2\left[1 + \sin^2\phi\right] \equiv \log_2\left[1 + p_e\right].
    \label{eqn_ln_sup_g_t_opt}      
\end{align}

\subsection{\label{sec:qbt_osc_ent} Qubit-oscillator entanglement}
Until now, we have only focused on the entanglement between the oscillators. Now, we will analyze the temporal evolution of the bipartite entanglement between the qubit and the oscillator(s). For convenience, we trace out the contributions of the second oscillator and focus on the joint density matrix of the qubit and the first oscillator. For initial thermal qubit and ground state oscillators, it has the following form when expressed in the bases $\ket{g,\,0}$, $\ket{e,\,0}$, $\ket{g,\,m}$, and $\ket{e,\,m}$: 
\begin{align} 
    \rho^{\text{th}}_{o1,qb}(t) = 
    \begin{pmatrix}
    |x_1|^2 + |x_4|^2 & 0 & 0 & 0 \\
    0 & |x_2|^2 & x_2x_3^* & 0 \\
    0 & x_3x_2^* & |x_3|^2 & 0 \\
    0 & 0 & 0 & 0
    \end{pmatrix}. 
    \label{rho_qo}
\end{align}
Note the similarity between Eq.\eqref{eqn_rho_12_q_th} and Eq.\eqref{rho_qo} under the exchange of the coefficients $x_2\leftrightarrow x_4$. 
Following a similar analysis, we find that only one eigenvalue can become negative. It is given by
\begin{align} 
  \lambda_- = \frac{1}{2} \left[|x_1|^2+|x_4|^2 - \sqrt{\left(|x_1|^2+|x_4|^2\right)^2 + 4|x_2|^2 |x_3|^2} \right].  
  \label{eq35b}
\end{align}

Therefore, the logarithmic negativity $\mathcal{L}^{\text{qo}}$ for the qubit oscillator subsystem is given by
\begin{align} 
    \mathcal{L}^{\text{qo}} = 
    \log_2 \left[ 1 - \tfrac{\sin^2\tau}{2} + \sqrt{\tfrac{\sin^4\tau}{4} + 2\sin^2\tau\cos^2\tau} \right].
    \label{eqn_l_qo}
\end{align}

On the other hand for the initial coherent spin state, we have
\begin{align} \label{c13}
    \rho^{\text{sup}}_{o1,qb}(t) = 
    \begin{pmatrix}
\vert x_1\vert^2 +\vert x_4\vert^2 & x_1 x_2^* & x_1 x_3^* & 0\\
x_1^* x_2 & \vert x_2\vert^2 & x_2x_3^* & 0\\
x_1^* x_3 & x_3x_2^* & \vert x_3\vert^2 & 0 \\
0 & 0 & 0 & 0
\end{pmatrix}.    
    \end{align}
Once again, by swapping $x_2$ with $x_4$ in Eq.\eqref{eqn_rho_12_q_sup}, we can obtain the partial transposed density matrix in Eq.~\eqref{c13}. 

 \begin{figure*}[ht!]
    \centering
    \includegraphics{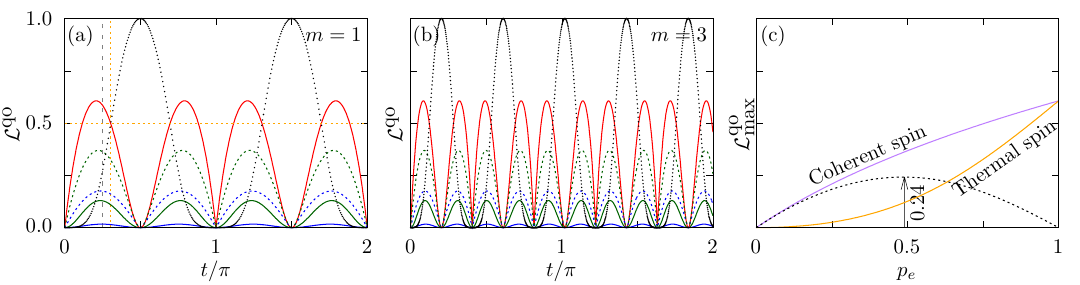}
    \vspace{-4ex}
    \caption{Similar to Fig.~\ref{fig_LN_Osc_vac} but instead of oscillator-oscillator entanglement, we show the behavior of the qubit-oscillator entanglement. The black dotted curves in panels (a) and (b) correspond to the oscillator-oscillator entanglement $\mathcal{L}$ for $p_e=1$. The vertical dashed line in panel (a) corresponds to $t=\pi/4$, while the yellow dashed lines in panel (a) show the interaction of the red curve and the black dotted curve.}
    \label{fig_LN_qub_osc}
\end{figure*}

Similar to the case with oscillator-oscillator entanglement, we were unable to write down an exact analytical expression for $\mathcal{L}^{\text{qo}}$ for this case.
In Fig.~\ref{fig_LN_qub_osc}, we compare the unitary evolution of $\mathcal{L}^{\text{qo}}$ for initial coherent (dashed curves) as well as incoherent (solid curves) spin state, setting $g_1=g_2=\frac{1}{\sqrt{2}}$ and $\Delta=0$. Note that for this case, $\vert x_1\vert^2=1-p_e$, $\vert x_2\vert^2=p_e\cos^2\tau$, and $\vert x_3\vert^2=\vert x_4\vert^2=\frac{1}{2}p_e\sin^2\tau$. 
Similar to the oscillator-oscillator entanglement $\mathcal{L}$ in Fig.~\ref{fig_LN_Osc_vac}, $\mathcal{L}^{\text{qo}}$ also exhibits oscillatory dynamics and the coherently prepared spin produces higher entanglement.

\subsection{\label{sec:short_time_ent} Differences in the behavior of $\mathcal{L}$ and $\mathcal{L}^{\text{qo}}$}
Comparing the behavior of $\mathcal{L}$ in Fig.~\ref{fig_LN_Osc_vac} and $\mathcal{L}^{\text{qo}}$ in Fig.~\ref{fig_LN_qub_osc}, we find three major qualitative differences:  (i) the entanglement in the qubit-oscillator subsystem $\mathcal{L}^{\text{qo}}$ materializes much faster than the oscillator-oscillator entanglement $\mathcal{L}$, (ii) the frequency of oscillations in $\mathcal{L}^{\text{qo}}$ is twice that of $\mathcal{L}$, and (iii) the maximum value of $\mathcal{L}^{\text{qo}}$ is only approximately 0.61, compared to the maximal value of $\mathcal{L}$, which is 1. We address them analytically in the following.





Observation (i) is intuitively expected, since the two oscillators do not interact directly but rather via the qubit. Consequently, it takes a while for the correlations between the oscillators to build up compared to those between the qubit and the oscillator. Mathematically, this can be explained by looking at their behavior near $\tau = 0$. The function $\mathcal{L}$ in Eq.~\eqref{fig_LN_qub_osc} for $p_e = 1$ near $\tau = 0$ behaves like $\mathcal{L} \approx \tau^4$, as shown below:
\begin{align} 
   \mathcal{L} &= \log_2\Big[\sin^2\tau  +\sqrt{1 - 2\sin^2\tau\cos^2\tau}\Big] \nonumber\\
   &\approx\log_2\Big[1 + \sin^4\tau\Big]\approx \sin^4\tau \sim \tau^4.
   \label{eqn_L_small_t}
\end{align}
On the other hand, the function $\mathcal{L}_\text{qo}$ in Eq.~\eqref{eqn_l_qo} for $p_e=1$ and $g_1=g_2$ near $\tau=0$ behaves linearly, that is, $\mathcal{L}_\text{qo} \approx \tau$ as shown below:
\begin{align} 
   \mathcal{L}^{\text{qo}} 
   &=\log_2 \left[ 1 - \tfrac{\sin^2\tau}{2} + \sqrt{2}\sin\tau\cos\tau\sqrt{1+\tfrac{1}{8}\tan^2\tau} \right] \nonumber\\
   &\approx\log_2 \left[ 1 + \tfrac{\sin(2\tau)}{\sqrt{2}} - \tfrac{\sin^2\tau}{2} 
 +\tfrac{1}{8\sqrt{2}}\tfrac{\sin^3\tau}{\cos\tau} \right]\nonumber\\
 &\approx \tfrac{1}{\sqrt{2}}\sin(2\tau) \sim \tau.
   \label{eqn_L_small_t_qo}
\end{align}

Furthermore, the fact that the frequency of the leading-order term of $\mathcal{L}^{\text{qo}}$ in Eq.\eqref{eqn_L_small_t_qo} is twice that of $\mathcal{L}$ in Eq.\eqref{eqn_L_small_t} explains the differences in the frequencies of oscillations for $\mathcal{L}$ and $\mathcal{L}^{\text{qo}}$. This also means that when $\mathcal{L}$ is maximum,  $\mathcal{L}^{\text{qo}}$ becomes identically zero, and when $\mathcal{L}^{\text{qo}}$ reaches its highest value, the oscillators are only weakly entangled. This behavior can be attributed to the monogamy of entanglement, which places constraints on the strength of the correlations between multiple parties. In essence, if some bipartite combination of a multipartite quantum system exhibits significant entanglement, then the remaining bipartite combinations are expected to generate less entanglement. 

Another interesting observation pertaining to the dynamics of $\mathcal{L}^{\text{qo}}$ is that the maximal $\mathcal{L}^{\text{qo}}$ is achieved not for $\tau=(2k+1)\pi/4$ but for $\approx 0.21(2k+1)\pi$ which we have obtained numerically, where $k=0,1,2,\cdots$. 
Substituting this optimal $\tau$ value in Eq.\eqref{eqn_l_qo}, we obtain
\begin{align}
\mathcal{L}^{\text{qo}}_{\text{max}} \approx \log_2 [1 + (0.53) p_e].
\end{align}
For $p_e=1$, we get $\mathcal{L}^{\text{qo}}_{\text{max}}\approx0.61$, which is much less compared to the unit value achievable for the oscillator-oscillator entanglement.



\begin{figure*}[ht!]
    \centering
    \includegraphics{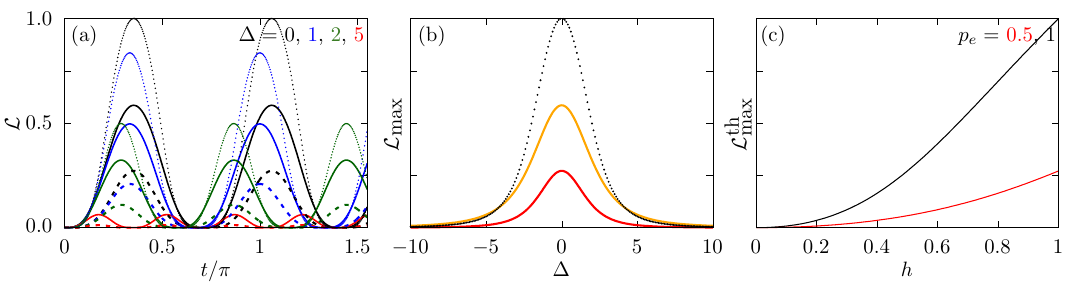}
    \vspace{-4ex}
    \caption{(a) The unitary temporal evolution of $\mathcal{L}$ is displayed in the presence of a common detuning $\Delta$, with different curves representing various values $\Delta$. The solid (dashed) lines correspond to the initial incoherent thermal (coherent) spin state with $p_e = 0.5$ while the dotted curves correspond to $p_e=1$. Both bosonic modes are initially in their respective ground states. Parameters are fixed at $m=1$ and $g=g_1 = g_2 = \frac{1}{\sqrt{2}}$ for consistency. Although the evolution maintains an oscillatory nature, the frequency of oscillations changes with the departure from perfect resonance ($\Delta=0$). More importantly, entanglement production is significantly diminished. (b) Illustration of the maximum achievable bosonic entanglement, $\mathcal{L}_{\text{max}}$, as a function of $\Delta$ for an initial incoherent (yellow) and a coherent (red) spin with $p_e=0.5$ while the black dotted curve corresponds to $p_e=1$. It is noteworthy that these two curves remain consistent regardless of the chosen values for $m$ and $g$. (c) $\mathcal{L}_{\text{max}}(h)$ in Eq.\eqref{eqn_LN_h} as a function of $h$.}
    \label{fig_LN_Osc_vac_det}
\end{figure*}

\section{\label{sec:coupled_eqns} General solution and the role of detuning}
In this section, we present the general solution with a non-zero detuning $\Delta$ for the two-mode $m$-photon JC Hamiltonian as given in Eq.\eqref{eqn_mpjcm}. We consider the initial condition where the spin is in a generic superposition state, and the two bosonic modes are in their ground states, as described by Eq.\eqref{psi0_osc_vac}. For such an initial state, the state vector at a subsequent time exhibits a structure similar to that shown in Eq.\eqref{eqn_psi_full_sup_qbt_vac_osc}.

Now, using the Schr\"odinger equation we get four time-dependent coefficients that obey the following coupled differential equations
\begin{subequations}
\begin{align}
    i \dot{x}_1(t) &= -\frac{\omega_0}{2} x_1,  \\
    i \dot{x}_2(t) &= \frac{\omega_0}{2} x_2 + \sqrt{m!} g_1 x_3 +\sqrt{m!} g_2 x_4, \\
    i \dot{x}_3(t) &= \left(\frac{\omega_0}{2} -\Delta_1\right)x_3 +\sqrt{m!} g_1 x_2,   \\
    i \dot{x}_4(t) &= \left(\frac{\omega_0}{2} -\Delta_2\right) x_4 +\sqrt{m!} g_2 x_2.
\end{align}    
\label{eqn_x1_x4_dot}
\end{subequations}
Solving these equations in {\it Mathematica} with the initial conditions $x_1(0) = \cos\phi$, $x_2(0) = \sin\phi$, $x_3(0) = x_4(0) = 0$ and assuming $\Delta_1=\Delta_2=\Delta$ for simplicity, we obtain
\begin{subequations}
\begin{align}
    x_1 = &\cos\phi \,e^{i\omega_0 t/2}, \\
    x_2 = &  \sin\phi\, \Big[\cos\left(\tfrac{1}{2}\sqrt{4 m! \tilde{g}^2 + \Delta^2}\, t\right) - i\frac{\Delta }{\sqrt{4 m!\tilde{g}^2 + \Delta^2}} \nonumber\\
    &\quad\quad\times \sin\left(\tfrac{1}{2}\sqrt{4 m! \tilde{g}^2 + \Delta^2}\, t\right)\Big] e^{\frac{i}{2}(\Delta - \omega_0)t},  \\
    x_3 = &  -i\sqrt{\frac{4 m! g_1^2}{4 m! \tilde{g}^2 + \Delta^2}}  \sin\phi \, \sin\left(\tfrac{1}{2}\sqrt{4 m! \tilde{g}^2 + \Delta^2}\, t\right)  \nonumber \\
    &\quad\quad\quad\times  e^{\frac{i}{2}(\Delta - \omega_0)t}, \\
    x_4 = &  -i\sqrt{\frac{4 m! g_2^2}{4 m! \tilde{g}^2+ \Delta^2}}\sin\phi \, \sin\left(\tfrac{1}{2}\sqrt{4m! \tilde{g}^2 + \Delta^2}\, t\right) \nonumber \\
    &\quad\quad\quad\times  e^{\frac{i}{2}(\Delta - \omega_0)t},
\end{align}
\label{eqn_x1_x4_delta}
\end{subequations}
where $\tilde{g}=\sqrt{g_1^2+g_2^2}$. The solution for $x_2$, $x_3$, and $x_4$ for the generic case when $\Delta_1\ne\Delta_2$, however, appears to be somewhat challenging even when using computational tools such as Mathematica. Note that the reduced two-mode bosonic density matrix at a later time $t$ retains the same structure as presented in Eq.\eqref{eqn_rho_12_q_sup}, albeit with modified coefficients given by Eq.\eqref{eqn_x1_x4_delta}. 

Similarly, it can be easily shown that for an incoherent noisy initial spin state, the expressions for the tripartite density matrix, the reduced two-mode bosonic density matrix, and the logarithmic negativity are governed by Eqs.\eqref{eqn_rho_full_th_qbt_vac_osc}, \eqref{eqn_rho_12_q_th}, and~\eqref{eqn_L_th_x1_4}, respectively, with time-dependent coefficients given by Eq.\eqref{eqn_x1_x4_delta}. In particular, the logarithmic negativity for the oscillator-oscillator subsystem can be expressed as 
\begin{align}
   \mathcal{L} &=\log_2 \Big[h\, p_e\sin^2\tau \nonumber\\
   &\quad\quad\quad+\sqrt{\left(1-h \,p_e\sin^2\tau\right)^2 +  g_{12}^2 h^2\, p_e^2\sin^4\tau}\Big],
\label{eqn_LN_delta}
\end{align}
where $\tau=\tfrac{1}{2}\sqrt{4m!\tilde{g}^2 + \Delta^2}\, t$ and $h = \frac{4m!\tilde{g}^2}{\Delta^2+4m!\tilde{g}^2}$. Clearly, $h\leqslant 1$ with the equality condition satisfying for $\Delta=0$. Also, the higher the value of $\Delta$, the lower the value of $h$. In particular, when $\sin\tau=1$, we obtain 
\begin{align}
   \mathcal{L}^{\text{th}}_{\text{max}}(h)= \log_2 \Big[h p_e +\sqrt{\left(1-h p_e\right)^2 + h^2p_e^2}\Big]. 
\label{eqn_LN_h}
\end{align}

In Fig.~\ref{fig_LN_Osc_vac_det}, the impact of imperfect frequency matching on entanglement is evident. It is apparent that even a slight deviation from perfect resonance results in diminished entanglement. Moreover, in the dispersive limit, specifically when  $\Delta\gg g$, the entanglement vanishes completely. This aligns with the anticipated dynamics of the dispersive regime~\cite{gerry_knight_2004}. In this regime ($g/\Delta\ll 1$), the Hamiltonian $H$ in Eq.\eqref{eqn_mpjcm} can be reformulated into an effective Hamiltonian $H_\text{eff}$, where the frequencies $\omega$ and $\omega_0$ undergo a constant shift, as discussed in~\cite{laha_josab_2023} for $m=1$. In Fig.~\ref{fig_LN_Osc_vac_det}(c) the function $\mathcal{L}^{\text{th}}_{\text{max}}(h)$ in Eq.\eqref{eqn_LN_h} is plotted in the range $0\leqslant h\leqslant1$. It increases monotonically with $h$ and attains the maximum value of unity when $h=1$, corresponding to $\Delta=0$.
\section{\label{sec:log_neg_gauss} Separability of reference Gaussian states}
In this Section, we show that the entanglement of the reference two-mode Gaussian state is zero at all times, for initial coherent as well as incoherent spin states with initial ground state bosonic modes. For this, we need to compute, from $\rho_{\text{b}}^{\text{sup}}(t)$ and $\rho_{\text{b}}^{\text{th}}(t)$, the corresponding mean vector $R = (x_a, p_a, x_b, p_b)$ and the real $4\times4$ covariance matrix $V$ with elements
\begin{equation}
    V_{ij} = \frac{1}{2}\langle R_i R_j + R_j R_i\rangle - \langle R_i \rangle\langle R_j\rangle.
\end{equation}
If we express the matrix $V$ in the standard block form as
\begin{align}
V =
\begin{pmatrix}
A & C\\
C^T & B
\end{pmatrix},
\label{cov_mat_standard}
\end{align}
where $A$, $B$, and $C$ are all real $2\times2$ matrices, the quantification of entanglement within the two-mode Gaussian state can be derived from the logarithmic negativity, as given by~\cite{serafini2017}
\begin{align}
\mathcal{L}_{\text{Gauss}} = \max\left\{0, -\frac{1}{2}\log_2\left[2 f\right]\right\},
\end{align}
where
\begin{align}
    f &= \det{A} + \det{B} - 2\det{C}  \nonumber\\
    &\quad- \sqrt{(\det{A} + \det{B} - 2\det{C})^2 - 4\det{V}}.
\end{align}

However, demonstrating the separability of the two-mode Gaussian state requires merely establishing $\det{C} \geqslant 0$~\cite{Simon_PRL_2000}.

For $\rho_{\text{b}}^{\text{sup}}(t)$ in Eq.\eqref{eqn_rho_12_q_sup} it is straightforward to show that
\begin{widetext}
\begin{subequations}
\begin{align}
 C_{11} &= \frac{1}{2} \sqrt{m}\, x_3 x_4^* \delta_{0,m-1} - \frac{1}{2} [\delta_{m1} (x_1x_3^*) + \sqrt{m} \delta_{0,m-1} (x_1^*x_3)] [\delta_{m1} (x_1x_4^*) + \sqrt{m} \delta_{0,m-1} (x_1^*x_4)], \\
 C_{12} &= \frac{-1}{2} \sqrt{m}\, x_3 x_4^* \delta_{0,m-1} - \frac{i}{2} [\delta_{m1} (x_1x_3^*) + \sqrt{m} \delta_{0,m-1} (x_1^*x_3)] [\delta_{m1} (x_1x_4^*) - \sqrt{m} \delta_{0,m-1} (x_1^*x_4)], \\
 C_{21} &= \frac{1}{2} \sqrt{m}\, x_3 x_4^* \delta_{0,m-1} - \frac{i}{2} [\delta_{m1} (x_1x_4^*) + \sqrt{m} \delta_{0,m-1} (x_1^*x_4)] [\delta_{m1} (x_1x_3^*) - \sqrt{m} \delta_{0,m-1} (x_1^*x_3)], \\
 C_{22} &= \frac{-1}{2} \sqrt{m}\, x_3 x_4^* \delta_{0,m-1} + \frac{1}{2} [\delta_{m1} (x_1x_3^*) - \sqrt{m} \delta_{0,m-1} (x_1^*x_3)] [\delta_{m1} (x_1x_4^*) - \sqrt{m} \delta_{0,m-1} (x_1^*x_4)].
\end{align}
\end{subequations}
\end{widetext}
Now, for $m\geqslant 2$, it is evident that $C$ is the null matrix implying that the states are always separable. For $m=1$, we have 
\begin{align} 
\det C = \frac{g_1^2g_2^2}{2 \sqrt{2}\tilde{g}^4} \sin^2 \phi \sin^2 (2\phi) \sin^4 (\tilde{g} t) \cos^2 (\omega_0 t+\frac{\pi}{4}).
\label{vmc50}
\end{align}
Because $\det C \geq 0$ for all $t$, the associated two-mode Gaussian states are always separable.

Similarly, for $\rho_{\text{b}}^{\text{th}}(t)$ in Eq.\eqref{eqn_rho_12_q_th}, we get
\begin{align}
C = \frac{\sqrt{m}\,\delta_{0,m-1}\,x_3 x_4^*}{2} 
\begin{bmatrix}
1 & -1 \\
1 & -1 
\end{bmatrix}.
    \label{cov_mat_th}
\end{align}
Evidently, $\det C=0$ for all $m \geq 1$.

\section{\label{sec:kerr_eqns} Details on Kerr nonlinearity}
In this Section, we write down the modified coupled differential equations that describe the state of the two-mode MPJC Hamiltonian with additional Kerr nonlinearities as given by $H_{\text{Kerr}}$ in Eq.\eqref{mpjcm_kerr} for the initial state given in Eq.\eqref{psi0_osc_vac} with $\Delta=0$.
These are
\begin{subequations}
\begin{align}
    i \dot{x}_1(t) &= -\frac{\omega_0}{2} x_1,  \\
    i \dot{x}_2(t) &= \frac{\omega_0}{2} x_2 + \sqrt{m!} \,(g_1 x_3 + g_2 x_4), \\
    i \dot{x}_3(t) &= \left(\frac{\omega_0}{2} + \chi_1(m^2-m)\right)x_3 +\sqrt{m!} g_1 x_2,   \\
    i \dot{x}_4(t) &= \left(\frac{\omega_0}{2} + \chi_2(m^2-m) \right ) x_4 +\sqrt{m!} g_2 x_2.
\end{align}    
\label{eqn_x1_x4_kerr}
\end{subequations}
It is noteworthy that Eq.\eqref{eqn_x1_x4_kerr} and Eq.\eqref{eqn_x1_x4_dot} are identical under the assumption  $\Delta_i = -\chi_i(m^2-m)$ (where $i=1,\,2$). Thus, the inferences drawn from Eq.\eqref{eqn_x1_x4_dot} hold true for this case as well. Specifically, the most generic solution for $x_2$, $x_3$, and $x_4$ remains elusive, although the solution for the symmetric case can be expressed exactly in terms of Eq.\eqref{eqn_x1_x4_delta} with the identification $\Delta=-\chi(m^2-m)$. In addition, we also note that for the standard JC interaction with $m=1$, the solution remains independent of the strength of the Kerr nonlinearity parameter $\chi$. 


\section{\label{sec:w_state} Hybrid tripartite entanglement}
In this Section, we go beyond the usual dynamics of bipartite entanglement and briefly analyze some interesting aspects of the hybrid tripartite entanglement. We have already noted that if the oscillators are initialized in their respective ground states, they can be {\em effectively} treated as qubits, as only $\ket{0}$ and $\ket{m}$ contribute to the unitary dynamics. In this restricted scenario, we ask the following question. By suitably adjusting the system parameters, is it possible to engineer meaningful tripartite entangled states (such as the W state $\frac{1}{\sqrt{3}}\left(\ket{1,\,0,\,0} +\ket{0,\,1,\,0}+\ket{0,\,0,\,1}\right)$~\cite{Dur2000} or the GHZ state: $\frac{1}{\sqrt{2}}\left(\ket{0,\,0,\,0} +\ket{1,\,1,\,1}\right)$)~\cite{ghz89} that have several applications in quantum information theory? We address this below assuming that the qubit is initially prepared in the excited state.

Apart from a global phase, we have already shown that the initial tripartite state $\ket{e,\,0,\,0}$ evolves under the MPJC Hamiltonian to the following state (assuming $g_1=g_2=g$)
\begin{align}
  \ket{\psi(t)} = x_2(t)\ket{e, 0, 0} + x_3(t) \ket{g, m, 0} + x_4(t) \ket{g, 0, m},
  \label{eqn_psit}
\end{align}
where $x_2(t)=\cos\tau$, $x_3(t)=x_4(t) = -\frac{i}{\sqrt{2}}\sin\tau$, where $\tau=\sqrt{2m!}\,gt$. 
Now, identifying $\ket{g}$ and $\ket{e}$ as $\ket{\bar{0}}$ and $\ket{\bar{1}}$ respectively, and the two effective oscillator states $\ket{0}$ and $\ket{m}$ as $\ket{\bar{0}}$ and $\ket{\bar{1}}$ respectively, we can rewrite Eq.\eqref{eqn_psit} as
\begin{align}
  \ket{\psi(t)} = x_2(t)\ket{\bar{1}, \bar{0}, \bar{0}} + x_3(t) \ket{\bar{0}, \bar{1}, \bar{0}} + x_4(t) \ket{\bar{0}, \bar{0}, \bar{1}}.
  \label{eqn_psit_w}
\end{align}
This is a genuinely tripartite quantum state as long as the three coefficients are nonzero. 

\begin{figure}[ht!]
    \centering
    \includegraphics{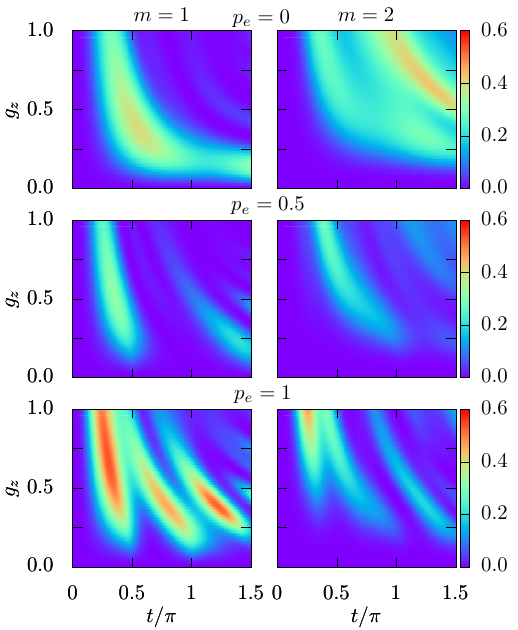}
    \vspace{-4ex}
    \caption{Emergence of spin coherence $\mathcal{C}$ arising exclusively from initially noisy and incoherent spins (with $p_e=0$, 0.5, and 1, respectively) within the single-mode $m$-photon Jaynes-Cummings model, augmented by an additional dispersive spin-boson coupling (see Eq.\eqref{eqn_mpjcm_1mode}). The dependence on time and the dispersive coupling parameter $g_{z}$ is illustrated. Assuming the initial ground states for the bosonic mode, and setting $g=1$, and $\Delta=0$ for consistency, the left and right columns represent cases for $m=1$ and 2, respectively. Similar to the two-mode case, here too coherence emerges even when $p_e=0$.}
    \label{fig_1mode_cq_Osc_vac_szx}
\end{figure}

Let us compare the red and the black dotted curves in Fig.~\ref{fig_LN_qub_osc}(a) which corresponds to $m=1$. We can identify two distinct aspects of bipartite entanglement during each cycle, apart from the trivial endpoints ($\tau=0$ and $\pi$ for the first cycle), where the tripartite state is completely separable. These are (i) $\tau\approx0.21\pi$ where the qubit-oscillator entanglement is maximum (as explained above) and where the two oscillators are very weakly entangled, and (ii) $\tau=\pi/2$ where the oscillators are maximally entangled but the qubit-oscillator entanglement is identically zero. However, between these two extreme cases, there is a time when all three bipartitions have equal entanglement. This is precisely the time when $\vert x_2\vert=\vert x_3\vert$. In other words, at the crossing of the red and black curves in Fig.~\ref{fig_LN_qub_osc}(a), $\cos\tau=\frac{1}{\sqrt{2}}\sin\tau$ (or $x_2(t)=\frac{1}{\sqrt{3}}$, $x_3(t)=x_4(t) = -\frac{i}{\sqrt{3}}$) so that we have
\begin{align}
  \ket{\psi(t)} = \tfrac{1}{\sqrt{3}}\left(\ket{\bar{1}, \bar{0}, \bar{0}} - i \ket{\bar{0}, \bar{1}, \bar{0}} -i \ket{\bar{0}, \bar{0}, \bar{1}}\right).
  \label{eqn_w_dynamics}
\end{align}
Evidently, apart from a global phase, $\ket{\psi(t)}$ in Eq.\eqref{eqn_w_dynamics} can be identified with the actual W state with a conditional phase rotation of the actual qubit.

On the other hand, it is straightforward to see that we can never engineer a GHZ-type tripartite entangled state at any point in the temporal evolution by simply manipulating the system parameters, even considering a generic superposition initial spin state.


\section{\label{sec:single_mode} Emergence of spin coherence from a bipartite spin-boson model}
In this Section, we examine the emergence of spin coherence from the noisy initial spin in a relatively simpler bipartite spin-boson model in which a single bosonic mode (described by boson creation and annihilation operators $a^\dagger$ and $a$, respectively) interacts with a single spin via the bipartite MPJC interaction with additional dispersive coupling. The Hamiltonian is given by
\begin{align}
    H =  \frac{\omega_0}{2}\sigma_z  + \omega a^{\dagger}a +  g\left(a^m\sigma_+ + a^{\dagger\, m} \sigma_-\right) + g_z \sigma_z X.
    \label{eqn_mpjcm_1mode}
\end{align}
Here, $g$ and $g_z$ correspond to the strength of the MPJC interaction and the dispersive interaction, respectively.
In Fig.~\ref{fig_1mode_cq_Osc_vac_szx}, we have shown the emergence of spin coherence $\mathcal{C}$ from the incoherent spin assuming that the bosonic mode is initialized in the ground state. Upon comparing Fig.~\ref{fig_cq_Osc_vac_szx} and  Fig.~\ref{fig_1mode_cq_Osc_vac_szx}, it is evident that the qualitative dependence of $\mathcal{C}$ on the system parameters is comparable.


\bibliography{reference}

\end{document}